\documentclass{aa}  
\usepackage{natbib}
\bibpunct{(}{)}{;}{a}{}{,} 
\usepackage[normalem]{ulem}
\usepackage{xcolor}
\usepackage{siunitx}
\usepackage[colorlinks=true,linkcolor=blue,citecolor=blue,urlcolor=blue]{hyperref}

\usepackage{hyperref}
\usepackage{graphicx}
\usepackage{txfonts}
\defcitealias{meidt24}{MvdW24}
\defcitealias{lynden-bell72}{LBK72}
\defcitealias{binney08}{BT08}

\begin{document} 

   \title{Transforming Galaxies with EASE: widespread structural changes enabled by short-lived spirals}
   \titlerunning{Transforming Galaxies with EASE}
   \author{
          Arjen van der Wel\inst{1}
          \and
          Sharon~E.~Meidt\inst{1}
          }

   \institute{Sterrenkundig Observatorium, Universiteit Gent, Krijgslaan 281 S9, 9000 Gent, Belgium}

 
  \abstract
    {
   We propose that galaxy structural changes -- and the rapid rise of a population of galaxies with early-type morphologies at cosmic noon ($1<z<3$) -- can be explained with EASE - Early, Accelerated, Secular Evolution. The mechanism relies on the torques exerted by stellar spirals in late-type galaxies that are present and active at $z>1.5$ as revealed by JWST/NIRCam.  The process is at once secular, because the transformative structural changes (heating, compaction, bulge formation) occur over many ($\approx10-30$) orbital periods, but accelerated, because orbital times were significantly shorter than at the present day.  
  In a first application, we take galaxy effective radius as a proxy for galaxy structure and, using new  measurements of the abundance and properties of stellar spirals observed in a collection of JWST deep fields, show that EASE predicts a distribution of early-type sizes that is smaller than late-type galaxies and consistent with what is observed.  
  The success of EASE relies on an updated picture of the influence of spiral arms, in which transience plays a key role.  We present a new calculation of the characteristic wave equation in the fluid approximation that applies to steady and non-steady open spirals beyond the more traditional tight-winding limit.  This shows open, transient spirals above the Jeans length growing and decaying on the order of a dynamical time in a wider region around and inside corotation than canonical steady spirals.  We show that this transient activity spreads out angular momentum gains and losses, and the associated dynamical heating, giving spirals a more extended influence than a single steady spiral. 
   The ubiquity of spirals in star-forming galaxies with stellar masses $M_{\star}>10^{10.5}~M_{\odot}$ 
   across the entire redshift range where early-type galaxies appear in large numbers suggests that EASE can play an important, or even dominant, role in morphological transformations across cosmic time.
  }

   \maketitle
%
%

\section{Introduction}\label{sec:intro}

One of the puzzles revealed by the observed evolution in the galaxy population over time is the need for a transformation process that is responsible — beyond a difference in initial conditions — for the distinct properties of the two main classes of galaxies, so-called early types and late types \citep{hubble26}.  The transformation process must evidently take the approximately exponential radial stellar light profiles of the younger late-type disk galaxies and help evolve it into the more peaked, concentrated and smoother stellar light distributions \citep{de-vaucouleurs59} of the older, less actively SF-ing early-type galaxies \citep{humason56, morgan57}, the majority of which retain rotation and disk like structure \citep[e.g.,][]{emsellem07, van-der-wel09a}. 

From a wealth of theoretical and observational studies over the last two decades, there are by now a number of mechanisms and evolutionary trajectories proposed to explain the variety in the properties throughout the observed galaxy population. 
Part of this variety can be expected to reflect differences in initial conditions \citep[e.g.,][]{van-den-bosch01}.   Old galaxies -- the early types -- formed when the Universe was younger and denser, leaving an imprint on present-day correlations between age, size, velocity dispersion and dynamical structure \citep{van-der-wel09, van-dokkum15, lilly16, chen20}. But the connection between the conditions during the main formation phase of the stellar body and the present-day state of a galaxy is easily scrambled by a number of additional processes.  Extended (star) formation histories, for one, can lead to the (re)growth of a disk at late times around a high-density center that formed at early times \citep{graham15}.  Then there are mergers that produce either classical (non-rotating) ellipticals or fast-rotating, disky early types \citep{bournaud04, naab06a, robertson06a, bournaud11}, depending on mass ratio and gas content.  In most galaxy formation models, mergers are the primary avenue toward bulge and spheroid formation \citep{cole00, somerville08, lacey16}. 

Another evolutionary channel is enabled by instabilities in disks. Central mass concentrations have been argued to arise with bar instabilities \citep{efstathiou82}, clump formation \citep{noguchi99, elmegreen04a, bournaud07a, elmegreen07, dekel09}, and violent disk instabilities  \citep{gammie01, krumholz10, dekel14}.  The wholesale transformation of the radial mass distribution brought about by disk instabilities in all of these cases is generically referred to as compaction \citep{dekel14, zolotov15, tacchella16, tacchella18}, with a distinction between processes that act `secularly', on many ($\gtrsim 10$) orbital time scales, like clump migration, and 'violently', on a single orbital time scale.

\begin{figure*}
    \centering
    \includegraphics[scale=.16]{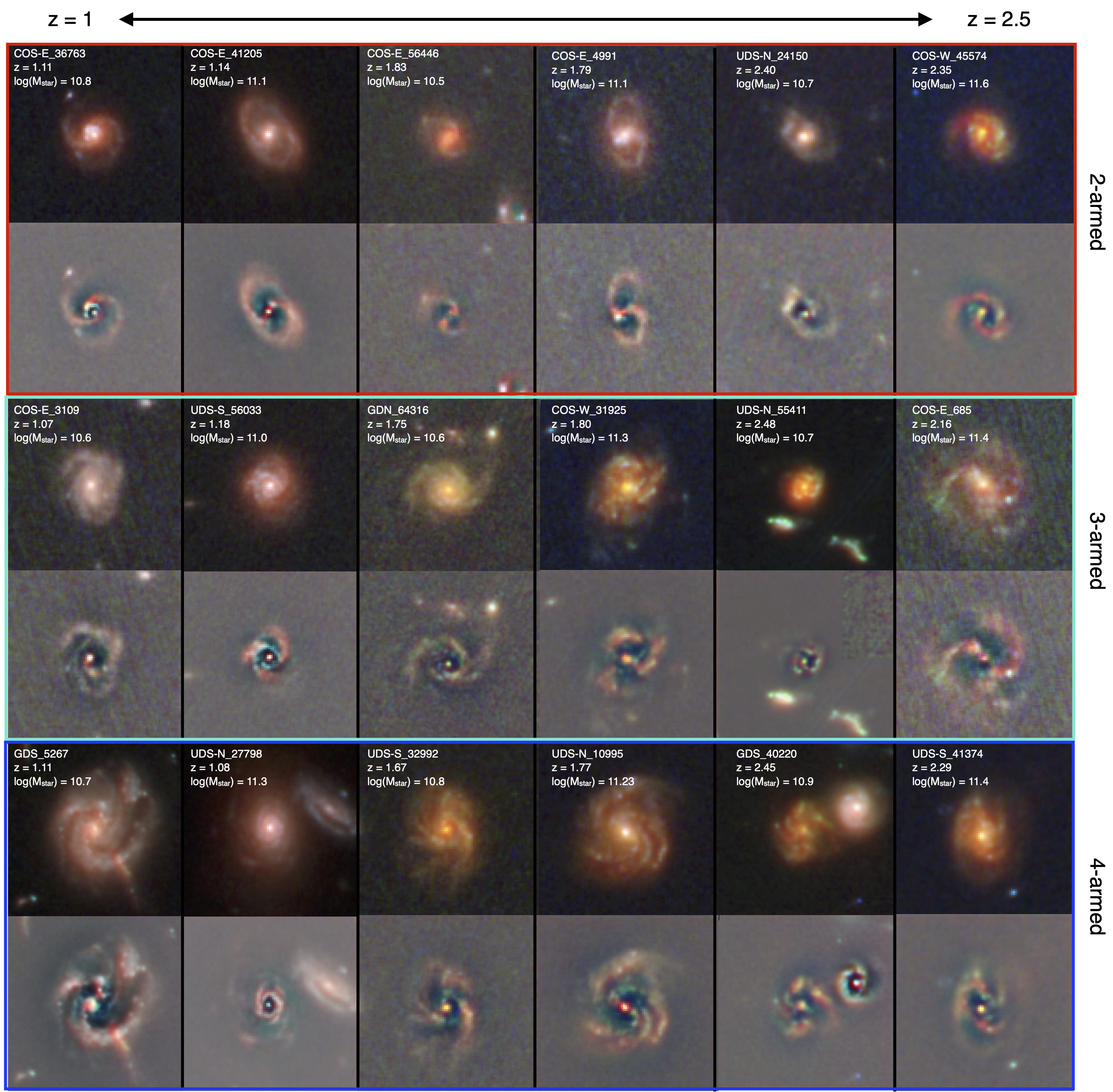}
    \caption{False-color three-filter JWST/NIRCam images of 18 spiral galaxies in the redshift range $1<z<2.5$ and in the stellar mass range $10.5<M_{\star}/M_{\odot}<11.5$. The images are 3.5 arcsec wide and high. Redshift increases from $z=1$ on the left to $z=2.5$ on the right, with two galaxies per redshift interval, one less massive than $M_{\star}=10^{11}~M_{\odot}$, and one more massive. For each galaxy we also show a residual image, highlighting the spiral structure, where a single S\'ersic component has been subtracted. The top row shows 2-armed spirals, the middle row shows 3-armed spirals, the bottom row shows 4-armed spirals. The number of arms is occasionally difficult to ascertain, but the presence of spiral structure is unambiguous in all 18 cases. The data and classification methodology are described in detail in Section \ref{sec:emp_data}. These 18 galaxies, drawn from the JADES and PRIMER NIRCam imaging surveys, are representative of the $> 500$ galaxies that were classified with high confidence as spirals. The IDs refer to the catalog entries in the DAWN  Arhcive catalogs derived from the various imaging mosaics (GOODS North/South, COSMOS East/West, UDS North/South). }
    \label{fig:images}
\end{figure*}

In this paper we aim to establish spiral arms as an early and powerful factor that is also capable of this kind of morphological transformation. This is a departure from the widespread expectation that spirals offer only a late-time influence, as they were expected to emerge only in significant numbers at late times ($z\lesssim 1$) after a more complex early history of disk building (at $z\gtrsim 1.5$).  JWST is forcing a change in this assessment, given the ubiquity of spiral structure in stellar disks at $z\sim 2$, as shown in Figure \ref{fig:images} and previously reported by \citet{guo23, ferreira23, kuhn24, espejo-salcedo25}.  Individual cases of galaxies with clear spiral structure have been discovered at even higher redshifts, $z=3-5$ \citep{costantin23, jain24, xiao25}.  The immediate implication of this empirical breakthrough is that the well-understood influence of spirals on their host galaxy's angular momentum and mass distributions, which have long been appreciated in the context of the local universe, are also important early factors in transforming galaxies along the Hubble sequence, aided by shorter dynamical time scales at earlier times (Appendix \ref{sec:Norb}). As we will argue in this paper, torquing by spiral arms may play a key role in the rise of the early-type galaxy population from $z=3$ to the present day. We call this Early, Accelerated Secular Evolution (EASE).

To build our view of the spiral influence at early times -- and establish spiral-driven secular evolution as a factor behind compaction --  in this work we take into account key elements that bring earlier accountings of the spiral influence up to date with the contemporary view of spirals as  transient, recurrent features \citep{goldreich65,julian66,toomre72, toomre81,sellwood84,sellwood89,sellwood19,sridhar19,sellwood22a}, that are capable of amplifying to prominence even in disks that are not as dynamically cold as in the local universe \citep{toomre81, meidt24}.  This helps us understand why spirals can be expected to form at such early times, when disks were hotter \citep{kassin07, law07, wisnioski15, ubler19}.  As we demonstrate with new calculations here, short-lived spirals moreover produce characteristically wider changes in angular momentum than predicted for canonical steady spirals.   So although any one transient feature may be short lived, a series of transient spirals can introduce not only significant changes in angular momentum, but also insure that the changes are radially extended.   

We begin in Section \ref{sec:spiral2} by giving a brief overview, aimed at a general audience, that describes the influence spirals are expected to have on galaxies. This is followed by a new description of the nature of transient spirals in Section \ref{sec:theory} and new calculations of the net torques exerted by spiral arms together with estimates of the radial redistribution of angular momentum in Section \ref{sec:ltrans}.  Sections \ref{sec:theory} and \ref{sec:ltrans} are aimed (not exclusively) at readers with a particular interest in disk dynamics. Then, in Section \ref{sec:emp}, we use JWST/NIRCam imaging data to measure the fraction of galaxies with spiral arms as a function of redshift (across the range $1<z<2.5$) and stellar mass (above $M_\star = 3\times 10^{10}~M_\odot$).   
Finally, in Section \ref{sec:appl} we will apply the theoretical calculations to the observed spiral galaxy population at redshifts $1<z<2.5$ to estimate the resulting changes in global galaxy structure (effective radius) with cosmic time.
Specifically, we examine the contribution of spiral arms torques to the `compaction' of star-forming galaxies at $z>1$ and the degree to which they can explain the significant difference between the radial stellar mass distributions of early- and late-type galaxies: early-type galaxies have smaller effective radii and more concentrated light profiles. In Section \ref{sec:discussion} we discuss our results in the context of existing work on disk instabilities and their effect on galaxy evolution.

\section{A brief overview of the spiral influence}
\label{sec:spiral2}
\subsection{Canonical Spiral torque estimation}
\label{sec:torque2}
At the foundation of the contemporary picture for how spiral arms influence their host galaxies is the calculation of exchanges of energy and angular momentum between the disk and the spiral at resonances, as first calculated by \citet[][hereafter, LBK72]{lynden-bell72} and \citet{goldreich79}.  The resonances occur when the spiral driving frequency is a multiple of the disk's natural orbital frequencies.  The pattern of resonant exchanges, combined with the facility of steady, trailing spirals in transporting angular momentum outward, make it possible for the galaxy to ultimately achieve a desirable lowest-energy, higher entropy state.  

In this picture, the changes brought about by any long-lived trailing spiral are predicted to be
significant 
\citep{gnedin95, binney08, foyle10}.  
The change in angular momentum due to a steady spiral is typically estimated from a straightforward calculation of the angular momentum current $C_{tot}$ that a small spiral potential perturbation carries across the disk.  This implicitly relies on the finding that the current, or wave action, for steady spirals is independent of radius \citep{goldreich79, toomre69}.  The idea is that whatever angular momentum the spiral liberated at corotation (through its growth) that eventually gets entirely removed at the inner Lindblad resonance (ILR; where it is absorbed) can be measured from the angular momentum current.

As given by \citet[][hereafter, BT08]{binney08}, in terms of the total angular momentum current $C_{tot}$ due to the sum of gravitational and advective torques\footnote{The advective torque results from the transportation (rather than gravitational potential) of disk material.}, the fractional change in the angular momentum inside radius $R$ after time $t$ can be estimated as
\begin{eqnarray}
\frac{\int C_{\rm tot}dt}{L=2\pi \int \Sigma_0 V_c R^2}&\approx&\textrm{sgn}(k)\frac{m V_c}{2 R}\frac{\Phi_a(0)^2}{V_c^4}\left(1-\frac{k}{k_J}\right)t\nonumber\\
&\approx& \textrm{sgn}(k)\frac{m}{2(kR)^2}\frac{\Sigma_a^2}{\Sigma_0^2}\frac{V_c}{R}\left(1-\frac{k}{k_J}\right)t\nonumber\\
&\approx& \textrm{sgn}(k)\frac{\pi \tan^2 i_p}{m}\frac{\Sigma_a^2}{\Sigma_0^2}N_{\rm rot}\left(1-\frac{k}{k_J}\right)
\label{eq:BT08Cg1}
\end{eqnarray} 
where for illustration a constant-velocity Mestel disk (an infinitely thin disk with $\Sigma\propto 1/R$ and flat rotation curve $V_c^2=2\pi G\Sigma R$) is adopted, as is the WKB approximation\footnote{The WKB approximation assumes that the phase variation of the perturbation is much greater than its amplitude variation, implemented in practice by taking $kR\gg1$, with $k$ the radial wavenumber of the perturbation.}, here impacting both the assumed relation between the perturbed density and potential $\Phi_a=-2\pi G\Sigma_a/\vert k\vert$ and the adopted expression for the advective torque (see \citealt{dehnen25} for alternative forms of the relation between the perturbed density and potential). The second of the two terms in parentheses arises with the advective torque and will tend to reduce the angular momentum transport achieved by spirals (\citetalias{lynden-bell72,binney08}).  Here $V_c$ is the circular velocity, $\Sigma_0$ is the unperturbed axisymmetric surface density at radius $R$, $N_{\rm rot}$ is the number of orbital periods $t_{\rm orb}=(2\pi R)/V_c$, $k_J=\sigma^2/(2\pi G\Sigma_0)$ is the inverse Jeans wavelength, $\Phi_a(0)$ is the magnitude of the potential at the mid-plane, $\Sigma_a$ is the perturbed surface density, $m$ is the spiral multiplicity, $k$ is the spiral wave number and $i_p=\arctan{m/(kR)}$ is the pitch angle. 

As discussed by \citetalias{binney08}, for reasonable, realistic spiral properties a steady spiral can produce a significant change to the distribution of the angular momentum in a galaxy disk after only a few orbital periods. 

\subsection{The timescales of spiral torques and the locations of their influence}

Ideally, Eq.~\ref{eq:BT08Cg1} alone would allow us to directly conclude whether or not spirals are a viable factor for transforming the morphologies of galaxies from late to early type.  When applying Eq.~\ref{eq:BT08Cg1} in practice, though, the net spiral impact is a sensitive question of time: when do spirals appear, do they evolve and how long do they last?    

For example, as highlighted by \cite{sellwood11}, the advective transport is minimized (and the total torque maximized) for open spirals, and specifically transient swing-amplified structures that are preferentially prominent near $k\rightarrow 0$.  This leads to the conclusion that the largest changes in angular momentum can be achieved by spirals that are open and non-steady, or transient, features.   

On the other hand, shearing material spiral patterns, which are expected even in \citet{toomre64} stable disks and thus under less favorable conditions than canonical spiral `modes', evolve so quickly that the briefness of their amplification is thought to be too strong a limit on torques and angular momentum changes to enact any lasting changes in isolation.  

The longevity and evolution of spirals is also important for where spirals exert their influence.  
 Even in disks where spirals can emerge relatively early and last for a significant number of dynamical times, the angular momentum changes (predicted by \citetalias{lynden-bell72}) occur exclusively at resonance, and thus over relatively narrow regions in the disk.  To have a more widespread impact, the spiral's pattern speed needs to evolve over time, so that resonances sweep out an extended zone in the disk.   It remains unclear how the torques exerted by steady spirals would be expected to lead specifically to the build up of the central mass concentrations that make early-types stand apart from their late-type counterparts.  This has also not been demonstrated for shearing, material spirals.

\subsection{Overview of the role of non-steady spirals in this work}
\label{sec:overview2}
In this work, we take a deeper look at how spiral longevity impacts the angular momentum distributions of their host disks. 
The transient nature of spirals is a solidifying element in our modern picture of how spirals originate and evolve \citep{sellwood84,sellwood19}, largely substantiated by numerical simulations \citep[e.g.][]{sellwood14,sellwood22}.  Those simulations have given us important clues as to how the transience emerges and have inspired recent analytical calculations aimed at describing how the transience originates, how it impacts the disk, and how it evolves \citep{sridhar19, hamilton24b,hamilton25}.  
In particular, the temporary action of a given spiral on the disk sets up other `groove mode' transient spirals, building an extended spiral patterns that extend far across the disk and stimulates new spirals continuously \citep{sellwood84,sellwood89,sellwood19}.  This is a promising manner to set up a widespread transport of angular momentum across the disk, over a longer period than the lifetime of any one transient spiral \citep{sellwood19,sellwood22}.

The particular aspect of transience that is the focus of this work is what it implies about wave action conservation and the potential for widespread change in angular momentum, whether the spirals are groove modes or shearing material patterns.  Our focus is on the basic idea that, since transience or non-steadiness alters the accounting behind the conservation of wave action, it can be expected to alter not just the amount that spirals impact the disk's angular momentum and structure (changing the total time the disk experiences a torque) but also where they exert their impact.  

To test this idea, in this paper we calculate the angular momentum current carried by a single transient spiral potential perturbation and use this to identify the scenarios where this can lead to mass flux through the disk.  
We also calculate the dynamical heating generated by both transient rigidly rotating modes and shearing patterns and explore how this heating depends on the growth rate of the transient spiral.

The torque calculations we present apply to any perturbation, in either gaseous or stellar disks, but while considering the expected temporal and spatial patterns of growth and decay along transient spirals, we have in mind the depiction of transience in recent linear hydrodynamical (fluid) calculations \citep[][hereafter, MvdW24]{meidt24} that we examine further here in the immediately following section. 
The hydrodynamical approximation is valid for spirals in stellar disks that can be treated in the linear regime and on scales above the Jeans length, which is our ultimate focus here.  The main value of the hydrodynamical approach is the greater ease with which the governing equations can be obtained and interpreted compared to collisionless, stellar-dynamical calculations. Like those calculations, our fluid approach is not sensitive to non-linear effects (see \citealt{hamilton25} for a review) and does not treat the result of evolution alongside other disk components \citep[e.g.,][]{jog84, bertin88}  

Our calculations can also be viewed as relevant for describing spirals in gas disks (which are interesting in their own right, though not the main focus of this work) and the radial gas inflows and dynamical heating they are predicted to generate.  Later in Section  \ref{sec:discussion} we discuss the relation between gaseous spirals and migrating clumps and the changes to the disk distribution that they bring about.  We note that factors like magnetic forces and star formation feedback are missing from the present approach, which must be thought of as revealing purely gravitational influences on gas and stars.

\section{The transient nature of spiral arms}
\label{sec:theory}

\subsection{Differences in Wave Action Conservation in Steady vs.~Non-Steady Spirals}

The changes in angular momentum brought about by spirals at resonances (\citetalias{lynden-bell72}) are intimately, but subtly, 
related to the conservation of wave action characteristic of steady spirals as they transport the angular momentum \citep{lin64,toomre69,goldreich79}. 
The amount of angular momentum carried by a steady spiral (the angular momentum current $C_{\rm tot}$; \citealt{goldreich79, binney08}) is independent of radius (unchanging along its length), making it 100\% efficient at transporting angular momentum between resonances.  In other words, the conservation of wave action in steady spirals insures that primary changes in angular momentum occur exclusively at narrow, resonant regions of the disk, and vice versa.  
How does this change when we weaken the requirement of steadiness?  

Wave action conservation is a feature of time-periodic wave solutions \citep{kato71, drury85}, i.e. those with fully real wave speeds $\omega$ that lack any growth or decay behavior.  This is the behavior assumed during the hydrodynamical calculation that implies wave action conservation for steady spirals \citep{goldreich78}.  Taking the wave equation in the WKB approximation or tight-winding limit, wave action conservation is obtained by collecting all the imaginary terms in the characteristic wave equation and requiring that they must necessarily cancel each other out, since the spiral is assumed to be steady.   
The potential for non-steadiness becomes more easily explicitly treated when, in contrast, imaginary terms are retained in the characteristic wave equation, as in the recent ``Bottom's dream'' approach adopted in \citetalias{meidt24} and designed for the purposes of going behind the tight-winding limit (to the regime of `open' spirals).  The intended result is a more general expression that applies in the open-spiral regime, well beyond the tight-winding limit of the Lin-Shu-Kalnajs dispersion relation.  Its greatest value is that it allows us to recognize and describe changes in wave action whether they are limited to resonance (as in the case of longer-lived spirals) or not, as we identify below. 

\subsection{Patterns of growth for Non-Steady Spirals}
The ``Bottom's Dream'' characteristic equation for open, non-steady spirals offers a powerful view of spiral growth and decay.  Direct access to the density and other disk quantities keeps the critical factors transparent and the result is an expression that is comparable to, but easier to obtain and interpret, than evolutionary equations in a purely collisionless stellar-dynamical approach.  As we highlight below, the equation contains a number of features that can be identified with behavior previously discussed in the literature related specifically to corotation amplification 
\citep[\citetalias{lynden-bell72},][]{drury85,kato03,kato16} and offers new insights as well.

For brevity we omit the full derivation (see \citetalias{meidt24}) and instead summarize the basic ingredients and assumptions of the calculation. 
Following many other hydrodynamical calculations, we combine the continuity equation
 \begin{equation}
\frac{\partial\rho}{\partial t}=\mathbf{\nabla}\cdot(\rho\mathbf{v})=0
\end{equation}
with Poisson's equation 
\begin{equation}
\nabla^2\Phi=4\pi G\rho\label{eq:poisson}
\end{equation}
and the solutions to the Euler equations of motion for the rotating disk plus a small perturbation, 
\begin{equation}
\frac{\partial\mathbf{v}}{\partial t}+(\mathbf{v}\cdot\mathbf{\nabla})\mathbf{v} = -\frac{1}{\rho}\mathbf\nabla p-\mathbf{\nabla}\Phi. \label{eq:EOM}
\end{equation}
Here $\rho$ is the gas density, $p$ is the thermal plus non-thermal gas pressure \citep[following][]{chandrasekhar51} and the gravitational potential $\Phi$ represents gas self-gravity together with a possible background potential defined by a surrounding distribution of gas, stars and dark matter.  The calculation is performed in a 3D, vertically extended disk.  

To satisfy the linearized perturbed equations of motion in cylindrical coordinates (identical to those in \cite{meidt22} and \citetalias{meidt24}), we adopt 3D perturbations of the form
\begin{equation}
\Phi_1(R,\phi,z,t)=Re[\mathcal{F}(R,z)e^{i(m\phi-\int\omega dt)}e^{ik R }e^{i k_z z}]  \label{eq:pertex}
\end{equation}

Here $\omega(t)=m\Omega_p+i\omega_i(t)$ is the complex, time-dependent oscillation frequency of the $m$-mode perturbation and the wavenumbers $k$=$2\pi/\lambda_r$ and $k_z$=$2\pi/\lambda_z$ describe the wavelengths of the perturbation in the radial and vertical directions, respectively \citep[e.g.][]{toomre64, goldreich65, lin64, binney08}.  Variations in the wave pattern speed are taken to be slow compared to the wave growth.

The calculation is performed entirely under the secular approximation, so all quantities in the disk are assumed to vary much more slowly in time than the perturbation.  We examine the evolution in disk properties as a result of the wave separately in Section \ref{sec:torques}.  
All wave properties and all quantities in the equilibrium disk are implicitly functions of galactocentric radius.  The radial variation in $k$ and in $m$ are taken to be weak, however, in order to resemble the shapes of observed spirals.  Thus the distances over which the instantaneous $m$ and $k$ vary are assumed to be considerably larger than the radial wavelength. (See \citetalias{meidt24} for a more thorough discussion of the assumptions.)  Since we allow for wave growth and decay and let $d\omega/dR\neq0$, the wave shape (orientation) can also be a function of time (see below).  We also focus on the regime of instantaneous or fast growth in which the spiral's speed and its evolution are much larger than any changes in shape, i.e. that $\omega\gg\dot{k}/k$.  

The resulting expression in its most general form is the following cubic equation:
\begin{eqnarray}
(\omega_e-m\Omega)^3&=&\label{eq:full3drelation}\nonumber\\
 \Bigg[\kappa^2&+&\left(k_e^2+\frac{m^2}{R^2}-ik_e\left(\frac{\partial \ln \rho_0}{\partial \ln R}+\frac{1}{R}-\frac{1}{R}\frac{\partial\ln\Delta}{\partial \ln R}\right)\right)s_0^2\nonumber\\
 &+&C_z \Delta\Bigg](\omega_e-m\Omega)\nonumber\\
&+&i s_0^2k_e\Bigg(\frac{2Am}{R}-\frac{\partial(\omega_e-m\Omega)}{\partial R}\Bigg)\\
&+&s_0^2\frac{m\Omega}{R}\left(\frac{\partial \ln \rho_0}{\partial \ln R}+\frac{\partial \ln \Omega}{\partial \ln R}-\frac{\partial \ln\Delta}{\partial \ln R}\right)\nonumber
\end{eqnarray}
where 
\begin{eqnarray}
\Omega&=&V_c/R\nonumber\\ 
B&=&-\Omega-\frac{1}{2}R\frac{d\Omega}{dR}\\\nonumber
A&=&-\frac{1}{2}R\frac{d\Omega}{dR}=\Omega+B\nonumber\\ 
\Delta&=&\kappa^2-(m\Omega-\omega_e)^2\\\nonumber
\kappa^2&=&-4B\Omega.\nonumber
\end{eqnarray}
and
\begin{equation}
k_e=k-\int\frac{\partial\omega}{\partial R}dt,
\end{equation}
\begin{equation}
\omega_e=\omega-R\frac{\partial k}{\partial t},
\end{equation}
\begin{equation}
s_0^2=\frac{\rho_0}{\rho_a}\chi_a, 
\end{equation}
\begin{equation}
\chi_a=\left(\Phi_a+\sigma^2\frac{\rho_a}{\rho_0}\right)
\end{equation}
in terms of an isotropic isothermal velocity dispersion $\sigma$.  Here 
\begin{equation}
\Phi_a=\frac{4\pi G\rho_a}{(ik_e+T_{R,0})^2+\frac{(ik_e+T_{R,0})}{R}-\frac{m^2}{R^2}+(ik_z+T_z)^2}\label{eq:poissonFull}
\end{equation}
according to Poisson's equation, 
where $T_R=dln\mathcal{F}(R,z)/dr=T_{R,0}-\int\frac{\partial\omega}{\partial R}dt$ and $T_z=dln\mathcal{F}(R,z)/dz$ represent the characteristic scales of the perturbation's amplitude variations in the radial and vertical directions. Here $T_{R,0}$ is meant to represent any initial radial amplitude variation. 
The factor $C_z$ in Eq.~\ref{eq:full3drelation} includes the vertical terms that are assumed to be negligible for the fastest growing spirals (see \citetalias{meidt24}).  

Note that, with the exception of the definition in Eq.~\ref{eq:poissonFull} all factors involving $T_r$ in Eq.~\ref{eq:full3drelation}, or an equivalent factor for quantifying the density perturbations amplitude variation, are left implicit for illustration.  These factors together represent the 'phase-shift' between the density and potential emphasized in other work \citep{zhang96, dehnen25}.  At any time we assume that the amplitude variation is mostly the result of differential wave amplification and many of the relevant insight about growth we wish to highlight here can be obtained at the stage that those factors are small.   We do implicitly take them into account as an indication that eventually growth shuts off once $T_r$ grows beyond $k$ (through the contribution $\int \partial\omega/\partial R dt$)  We note, though, that keeping those factors in is relevant for describing the ultimate cessation of growth. 

Eq.~\ref{eq:full3drelation} is a cubic relation that contains two sets of dominant terms.  The terms in square brackets represents the conventional stability regime and dominates in the tight-winding limit ($k\gg m/R$) and wherever $Q<1$.  In these cases, the cubic relation becomes the standard Lin-Shu-Kalnajs quadratic dispersion relation.  In the regime, $1 \lesssim Q\lesssim 6$, the constant term dominates the solutions, and we will focus on this regime in what follows.  Towards galaxy centers where $\kappa^2$ increases rapidly, the linear term becomes relevant and typically signifies that wave propagation is inhibited, except in special circumstances that are not the focus of the present work. 
As discussed in \citetalias{meidt24}, the cubic dispersion relation has complex factors, directly pointing to the possibility of growth and decay in each of the three possible solutions. 

\subsection{Resonant growth of open spirals in the short-wave limit}\label{sec:shortwaves}
In \citetalias{meidt24} we examined spiral growth specifically in the regime of open spirals with $kR\sim m$ in the short-wave (WKB) limit with $kR\gg 1$. In this case, the characteristic equation becomes
\begin{equation}
(\omega_e-m\Omega)^3\approx i\left(\frac{2Ak_em}{R}\right)s_0^2\label{eq:omegacross}
\end{equation}
where 
\begin{equation}
s_0^2=\frac{-4\pi G\rho_0}{k_t^2}+\sigma^2 \label{eq:s0sq}
\end{equation}
with
 \begin{eqnarray}
 k_t^2&=& k_e^2+\frac{m^2}{R^2}-T_z^2 \nonumber\\
&\approx& k_e^2+\frac{m^2}{R^2}.\label{eq:kteq}
\end{eqnarray}
 in the WKB approximation.  The final line of Eq.~\ref{eq:kteq} assumes that the vertical factor $T_z$ is negligible, as would be most favorable to instability as discussed in \citet{meidt22}.

One of the three solutions to the cubic characteristic equation in the short-wave regime is fully imaginary, corresponding to growth tightly localized around corotation.  Thus the characteristic equation is able to describe both the amplification of rigid spirals at corotation above $Q=1$ and the swing amplification of shearing material patterns, both of which we argue ultimately rely on the donkey effect\footnote{ Energy conservation and the characteristics of orbits in self-gravitating disks produce a push felt by perturbed stars at corotation that is directed away from (rather than toward) the perturbed (spiral) potential minimum. This was dubbed the donkey effect by Lynden-Bell.} and last only temporarily.\footnote{In the case of rigidly rotating waves, the growth eventually becomes so successful that wave's gravitational forcing no longer elicits the response needed to engage the donkey behavior (inverse Landau damping) responsible for wave growth.  In the case of shearing spirals, shear rotates the wave away from the critical orientation for optimized donkey behavior and eventually far enough so that self-gravity weakens relative to pressure, shutting off growth.  } 
In other words, with a single expression that can treat resonant and non-resonant zones, we can see changes in wave action in the short-wave limit that are isolated at resonance (as expected).  

\subsection{Resonant and non-resonant growth of open spirals in the long-wave limit}\label{sec:longwaves}
The long-wave regime $kR\approx 1$ becomes relevant as disks become warmer and the Jeans length increases, necessarily shifting the possibility of growth to larger scales.  (The short-wave regime remains most relevant in dynamically cold disks).   
In this case, 
\begin{eqnarray}
(\omega_e-m\Omega)^3&\approx&\label{eq:long}\\
&+&i s_0^2k_e\Bigg(\frac{2Am}{R}-\frac{\partial(\omega_e-m\Omega)}{\partial R}\Bigg)\nonumber\\
&+&s_0^2\frac{m\Omega}{R}\Bigg[\frac{\partial \ln \rho_0}{\partial \ln R}+\frac{\partial \ln \Omega}{\partial \ln R}-\frac{\partial \ln\Delta}{\partial \ln R}\Bigg]\nonumber
\end{eqnarray}
where \begin{equation}
s_0^2=4\pi G\rho_0\left(\frac{1}{k_{long}^2}+\frac{1}{k_J^2}\right) \label{eq:s0sqlong}
\end{equation}
in terms of $k_J=(4\pi G\rho_0/\sigma^2)$ and
 \begin{eqnarray}
 k_{long}^2&=(ik_e+T_{R,0})^2+\frac{(ik_e+T_{R,0})}{R}-\frac{m^2}{R^2}+T_z^2.
\end{eqnarray}
In this long-wave regime, terms scaling like $1/R$ in the characteristic cubic equation that were negligible in the short-wave limit now have powerful influence.  The result, as described below, is that growth broadens around corotation.  

We will illustrate this here by focusing on two terms, $d(\omega_e-m\Omega)/dR$ and $d\Delta/dR$, that are especially influential.  
Let us consider first the term $d(\omega_e-m\Omega)/dR$ and when it goes to zero, such as at or near corotation for a rigid wave.  Then the terms in parentheses in Eq.~\ref{eq:long} becomes $2Am/R$, while at all other locations this becomes i$\partial \omega_i/\partial R$.  Likewise, for material spirals, the term in parentheses is $2Am/R+i\partial \omega_i/\partial R$ everywhere.  This is a sign that, as in the short-wave limit, corotating long-wave spirals (or more generally where $d(\omega_e-m\Omega)/dR$=0) will be subject to growth.  The factor i$\partial \omega_i/\partial R$ implies that spirals are also capable of growth when and where they grow (or decay) by independent means, such as through an addition of mass added by star formation \citep[e.g][]{sellwood84}, or under the influence of gas, for example, but this avenue is not the focus here and will be neglected in what follows.  

Next consider the term $Rd\ln\Delta/dR$, which is one of three factors that make up the 'long-wave term' in the square brackets in Eq.~\ref{eq:long}.
Specifically at corotation (or more generally where $d(\omega_e-m\Omega)/dR$=0) this turns into $2d\ln\kappa/dR\approx 2(\beta-1)$ in terms of the logarithmic derivative of the rotation curve $\beta=d\ln V_c/d\ln R$.  Near to corotation, this becomes a factor 
$2(\beta-1)-\epsilon$ where $\epsilon$ is a small factor that increases with distance from the resonance. Very far from corotation, nearing the Lindblad resonances, $d\Delta/dR\rightarrow 0$.   It is straightforward to confirm that no complex solution above the Jeans scale exists specifically at the Lindblad resonances.    

Now rewriting the long-wave term in square brackets in Eq.~\ref{eq:long} at and near corotation, we have 
\begin{eqnarray}
\label{eq:longwaveterms}
\frac{d\ln\rho_0}{d\ln R}+\frac{d\ln\Omega}{d\ln R}-2\frac{d\ln\kappa}{d\ln R}&=&\epsilon-\frac{d\ln\kappa^2/(\Omega\rho_0)}{d\ln R}\\
&\approx&\epsilon+\frac{d\ln \mathcal{L}}{d\ln R}-2-2(\beta-1)\nonumber\\
&\approx&\epsilon+\frac{d\ln \mathcal{L}}{d\ln R}-2-\frac{\kappa^2-4\Omega^2}{2\Omega^2}\nonumber
\end{eqnarray}
where the angular momentum per unit area $\mathcal{L}=\Sigma\Omega R^2$ and the small positive factor $\epsilon$ is zero at corotation and increasingly large moving away from corotation. The first line highlights the equivalence of the long-wave factors to the specific vorticity featured in the context of corotation amplification \citep[][discussed in Section \ref{sec:waser}]{drury85,papaloizou89}.  The last two lines hold for flat rotation curves and point to the importance of the donkey effect, as we discuss in Appendix \ref{sec:donkey}.  

Altogether, combining the long-wave term with the imaginary factor present when $d(\omega_e-m\Omega)/dR\rightarrow 0$, the cubic dispersion relation admits only fully complex solutions. Two of these represent decay well inside or outside corotation.  The third contains a small real component that signifies growth shifted to just inside or outside corotation depending on the sign of the long-wave term.  When the long-wave term is positive, growth shifts to outside corotation (the real part of $\omega_e-m\Omega$ is positive), while growth shifts inside corotation when the long-wave term is negative.  This third growing solution approaches a purely imaginary solution that resembles corotation growth in the short-wave limit only as long as the long-wave terms remain small.  But as the long-wave term increases, we have a scenario in which the growth possible when $d(\omega_e-m\Omega)/dR\rightarrow 0$ shifts further and further away from corotation.  

Shifting outwards from corotation to the region where only the long-wave term contributes, one of the trio of solutions is fully real just outside (inside) corotation depending on whether the long-wave term is negative (positive).  The other two correspond either to growth or decay on the opposite side of corotation.  This has several notable consequences.  
In galactic disks the long-wave term will be almost exclusively negative (with a few exceptions), given the way $\Sigma$ and $\Omega$ typically both fall off with radius. Thus the fully stable region will sit outside corotation, while preferentially inside corotation is where both growing and decaying waves exist together.  
In all, spirals growing through the long-wave term are expected to be positioned inside corotation and persist even moving away from resonance, alongside spirals that were previously amplified but presently undergoing decay.   
Far enough from corotation, $d\Delta/dR$ deviates progressively more from $d\kappa^2/dR$ and $\epsilon$ in Eq.~\ref{eq:longwaveterms} increases, eventually making long-wave term positive and supressing growth.  

\subsubsection{Relation to corotation amplification and the WASER mechanism}\label{sec:waser}
As highlighted by the first line of Eq.~\ref{eq:longwaveterms}, the long-wave term is equivalent to the specific vorticity term $d(\kappa^2/(2\Omega\rho_0))/dR$ pivotal to corotation amplification in the calculations of \citet{drury85} and \citet{papaloizou89} \citep[see also][]{kato16}.  (\citet{papaloizou89} call the specific vorticity the vortensity.) Those hydrodynamical calculations reproduce the results of the series of collisionless calculations performed in \citet{mark74} and \citet{mark76}, in which corotation amplification is first cast in terms of the WASER mechanism that trades strength between incoming and outgoing waves.  
Our three solutions in the regime $d(\omega_e-m\Omega)/dR\rightarrow 0$ can be viewed as the WASER mechanism's incident wave at corotation together with the transmitted wave and the over-reflected wave that falls on either side of corotation depending on the sign of the specific vorticity.  This corotation amplification is also analogous to the Papaloizou-Pringle instability in thick gaseous torii \citep{ papaloizou84,kato16}.

In the long-wave characteristic equation in Eq.~\ref{eq:long}, the long-wave or specific vorticity term is present and influential even away from corotation.  In this light, a more straightforward and general way to view the growth and propagation of waves at and near corotation takes the perspective adopted by \citetalias{lynden-bell72}, which relates corotation amplification to the donkey effect, as we examine in  Appendix \ref{sec:donkey}.  This perspective is especially useful for understanding the near-resonant growth that occurs around corotation. 

\subsubsection{Differences with respect to the adiabatic approximation in \citetalias{lynden-bell72}}
In Appendix \ref{sec:donkey} we discuss the importance of the donkey effect for amplification in an extended island around corotation. 
The role of the donkey effect in the libration of stars in an extended corotation island is well known, as is its critical role in the horse-shoe behavior that gives rise to angular momentum changes during radial migration \citep{sellwood02}.  Those changes are present when the spiral lifetime is long enough that libration begins, but short enough that orbiting stars get stranded before completing a full libration.  

Even though the donkey effect is also pivotal to \citetalias{lynden-bell72} description of spiral amplification at corotation, it is worth noting that its importance particularly for the near-resonant growth of open, long-wave spirals evident here was not anticipated by the original calculation.  

The resonant changes in angular momentum revealed by the \citetalias{lynden-bell72} torque calculation rely on the adiabatic approximation, in which the spiral is assumed to be introduced slowly to the system.  This makes the derived angular momentum changes most sensitive to the longer-lived waves that remain only after all transient features have died away.  The \citetalias{lynden-bell72} torque calculation thus misses both transient, swing-amplifying material arms and the transient near-resonant growth found in this work.  The relevance of non-adiabatic, transient structures missed by the \citetalias{lynden-bell72} torque calculation has also been emphasized recently by \citet{banik23} and an earlier calculation by \citet{weinberg04} in the context of dynamical friction during disk-satellite interactions.  In line with our findings, those authors conclude that transience can be an influential factor for the secular structural evolution of galaxies. 

The non-resonant growth found here in the long-wave limit is worth considering in light of the similarity pointed out by \citet{polyachenko19} between their solutions to their general dispersion relation in the tight-winding limit and the \citetalias{lynden-bell72} torque formula.  The implication is that tight-winding spirals dominate the growth of steady spirals at corotation.  In other words, the limit $\gamma\rightarrow 0$ adopted by \citetalias{lynden-bell72} coincides with the tight-winding limit.  This may not be surprising given that the conservation of wave action away from resonance also holds in the tight winding limit \citep{goldreich79}.  But certainly, by extension it highlights that open spirals are either non-steady or they are important away from corotation, or both and, if transient, they would be missed by the adiabatic approximation.  

Still, given the importance of the donkey effect for the amplification identified by \citetalias{lynden-bell72}, we might have anticipated that wherever the donkey effect remains influential -- in the wider corotation island -- that spirals will be able to amplify, if only briefly.  
This idea already seems to be implicitly a part of how we envision spirals, given that -- based as much on how they appear in simulations as well as observations -- we accept that they are typically extended, broad features, rather than features that are prominent exceptionally in narrow zones.  The description provided by ``Bottom's Dream'' characteristic equation solidifies this view, not only implicating the donkey effect as much for resonant growth as for broader, near-resonant growth, 
but also showing that the amplified features near resonance are intrinsically transient, contrasting with the steady, long-lived modes predicted by \citetalias{lynden-bell72}.

\subsubsection{The growth rates and lifetimes of transient spiral patterns}\label{sec:lifetimes}
According to Eq.~\ref{eq:long}, the growth timescales for near-resonantly amplifying spirals will be on the order of a dynamical time at their fastest.  

At and near corotation the growth rate is approximately 
\begin{eqnarray}
\omega_{\mathrm{grow}}\approx \Bigg(2\Omega 4\pi G \rho_0\frac{m}{R^2}&&\left[\frac{1}{-k_{\rm long}^2}-\frac{1}{k_J^2}\right]\times\nonumber\\
&&\left[\frac{2d\ln\kappa}{d\ln R}-\frac{d\ln\Sigma\Omega}{d\ln R}\right]\Bigg)^{1/3}\label{eq:growratelong}
\end{eqnarray}
where the factors in the second set of square brackets are positive for growth. 

For open spirals well above the Jeans length $k\sim m/R<k_J$,  the fastest growth will satisfy
\begin{equation}
\omega_{\mathrm{grow}}\approx \Bigg(2\Omega 4\pi G \rho_0\left[\frac{2d\ln\kappa}{d\ln R}-\frac{d\ln\Sigma\Omega}{d\ln R}\right]\Bigg)^{1/3}.
\end{equation}
(The spiral properties $k$ and $m$, or pitch angle, for fastest growth can be obtained from Eq.~\ref{eq:growratelong}, but this is beyond the scope of this work.)

In self-gravitating power-law disks with $\Phi\propto R^{-\alpha}$ for $\alpha>0$ or $\Phi\propto \log(R)$ for $\alpha=0$, 
the term in square brackets becomes $3\alpha/2$ (see next section) and goes to zero specifically in the case of the Mestel disk.  When the disk represents only a fraction of the total mass of the galaxy such that the galaxy rotational velocity is decoupled from the disk's surface density, on the other hand, then the term in square brackets is very likely to be non-negligible. For an exponential density distribution $d\ln\Sigma/d\ln R=-R/R_d$, for example, 
\begin{equation}
\omega_{\mathrm{grow}}\approx \Bigg(2\Omega 4\pi G \rho_0\left[(\beta-1)+\frac{R}{R_d}\right]\Bigg)^{1/3}\label{eq:growthrate}
\end{equation}
now also letting $d \ln \kappa^2/d\ln R=2(\beta-1)$ assuming that $d\ln \beta/d\ln R\ll\beta$, as would be the case for a nearly flat rotation curve with $\beta\rightarrow 0$. 

Rewriting the prefix as $\kappa (4/Q_M)^{1/3}$ where $Q_M=\kappa^2/(4\pi G\rho_0)\approx 2Q_T$ in terms of the Toomre Q parameter \citep{meidt22}, then at fastest $\omega_{\rm grow}\approx\kappa$ between 1 and 2 disk scale lengths.  

With growth on the order of a dynamical time predicted in otherwise smooth disks, the donkey behavior it relies on eventually weakens, turning the growth to decay (see \citetalias{meidt24} and the discussion in Section \ref{sec:theory}) and finally shutting it off completely.  Eq.~(\ref{eq:growratelong}) suggests that the decay happens at a similar rate as the preceding growth, making the total lifetime of the transient spiral roughly a dynamical time.   

\subsection{Grooves, edges and impedance changes}
The long-wave term in Eq.~\ref{eq:long}, and the growth rate in eq. \ref{eq:growthrate}, highlight that local rapid changes in the disk density profile are sites where even more rapid growth can be stimulated.  This includes density gradients in the form of a `groove' or an edge (or outer disk truncation), for example, both of which have been shown in numerical simulations to sensitively influence wave growth and propagation \citep{sellwood89,sellwood91,sellwood19}.  
Locations with large local density gradients would behave as described in the previous section, launching waves inside or outside corotation depending on the sign of the density gradient, i.e. stimulating the 'groove modes' and 'edge modes' found in other analytical and numerical calculations \citep{toomre81, papaloizou89, sellwood89,sellwood91, de-rijcke16,sellwood19,binney20,fiteni24}. According to the convention discussed in the previous section, negative density gradients stimulate waves inside corotation, while an inner 'cut-out', or any positive gradient in the surface density, would stimulate a spiral outside corotation.  

\subsubsection{The ultimate fate of spiral patterns in disks}
\label{sec:fate}

With any given galaxy continuously exposed to a bath of perturbations at any given time\footnote{This includes earlier generations of stellar spirals, features in gas disks like molecular clouds and filaments, star formation feedback, and dark matter halo substructure, for example.}, broad, transient spiral patterns can be expected grow to prominence nearly continuously, provided disk conditions remain favorable.  
 
Expressing the factors in the long-wave term in terms of the angular momentum per unit area provides a powerful view of the conditions under which disks support spirals.  There are two features of note.  First, for reference, in the case of fully self-gravitating power-law disks with $\Phi\propto R^{-\alpha}$ for $\alpha>0$ or $\Phi\propto \log(R)$ for $\alpha=0$, the long-wave terms in Eq.~\ref{eq:longwaveterms} simplify to $-3\alpha/2$.  Corotation amplification is thus inhibited in Mestel disks where $\alpha=0$.  The ``Bottom's Dream'' equation -- and the essential role of the donkey effect in spiral amplification -- thus accounts for the well-known stability of the Mestel disk 
\citep[which only forms spirals through other means, like cut-outs, truncations, grooves and other impedance changes ][]{toomre81, zhang96, sellwood01, sellwood23}.

The second behavior of note is that spiral growth inside corotation occurs as long as the disk angular momentum continues increasing outward.  The factor $d\mathcal{L}/dR$ in the long-wave term in Eq.~\ref{eq:longwaveterms}
\begin{equation}
\frac{d\ln\mathcal{L}}{d\ln R}=\frac{R}{\mathcal{L}}\frac{d}{dR}\left(\frac{1}{R}\frac{dL}{dR}\right)=\frac{R}{\mathcal{L}}\left(-\frac{1}{R^2}\frac{d L}{dR}+\frac{1}{R}\frac{d^2 L}{dR^2}\right).
    \end{equation}
For a flat rotation curve with $\beta=0$, the long-wave term stays negative and corresponds to growth inside corotation as long as $d\ln\mathcal{L}/d\ln R$ stays negative or as long as
\begin{equation}
\frac{d L}{dR}>R\frac{d^2 L}{dR^2}.
\end{equation}

At the same time, spirals themselves act to transport angular momentum outward (\citetalias{lynden-bell72}; and see section \ref{sec:torques} below), if anything helping to steepen the rate of angular momentum increase.  This may eventually remove the need for a  spiral altogether, once the disk has achieved an ideal state with $d\mathcal{L}/dR\approx 0$.  In other words, the spiral is the disk's avenue for moving angular momentum outward up until the point that the outward transport brings the angular momentum to an optimum distribution.  The Mestel disk, for which  $d\mathcal{L}/dR$ is already zero, represents this ideal state, and thus no spiral is ever `needed' to transport angular momentum.  

The end result of the sensitivity of spiral formation to the angular momentum distribution is that spirals are eventually their own undoing.  This is also fundamentally behind the sensitivity of spiral formation and propagation to the dynamical hotness of disks, with disk heating viewed as inhibiting structure formation through a gradual elevation of the Toomre $Q$ parameter.  As demonstrated by \citetalias{lynden-bell72} and considered later in Section \ref{sec:heating}, dynamical heating is fundamentally the product of the outward transport of angular momentum.  The constraint on the angular momentum distribution as posed here may thus help to understand instances in which disks appear dynamically hot but nevertheless exhibit non-axisymmetric structures or when they are cold but surprisingly featureless. 

\subsection{Discussion}
The transient structures identified in this work resemble the dressed van Kampen modes identified in recent collisionless calculations that carefully treat wave solutions above and below the complex frequency plane \citep{lau21, lau21a}.  These calculations emphasize that transient waves, which are capable of growing when self-gravity is present in rotating systems, may be the more natural type of disk feature since no truly stable modes exist.
This work underscores that picture, showing with hydrodynamical calculations just how readily disks form transient features.

Implicit in our calculations and surrounding discussions is the remarkable similarity between rigidly-rotating near-resonantly excited transient spirals and shearing, swing-amplifying spiral patterns.  As discussed here and in \citetalias{meidt24}, both grow by relying on azimuthal forces prominent outside the tight-winding limit to engage \citetalias{lynden-bell72}'s donkey effect.  If transient rigid modes last for as long as shearing, swinging patterns remain prominent (on the order of a dynamical time), then the distinction between these two spiral types becomes especially blurry.  

Multiple, rigidly rotating spirals excited in serial, each at it's own corotation, end up tracing out a structure that behaves very much like an extended shearing pattern \citep{sellwood19, sellwood23}.  This suggest that, for estimating the impact of spirals on disks and their structural evolution, an extended, transient material spiral may provide as much insight as a more sophisticated model of overlapping modes.  

Rigidly rotating waves do have their advantages, especially for understanding recurrence.  As deciphered in some detail in the work of Sellwood \citep[e.g.][]{sellwood84,sellwood89,sellwood14,sellwood19,sellwood22}, ridigly rotating waves stimulate future sets of waves through the changes they introduce to the disk.  Those changes are absent from the present calculations in this section, but we make some effort in this direction by calculating the torques exerted by transient spirals and how they rearrange the angular momentum and mass distributions of galaxies in 
Section \ref{sec:torques}.
There we find that the strength and location with which a spiral impacts its underlying disk is very sensitive to whether the spiral is adiabatically or non-adiabatically (slowly or rapidly) growing.

\section{Angular Momentum Transport by Short-lived Spiral Arms}
\label{sec:ltrans}
\subsection{A heuristic model of Short-lived, Transient Spirals}\label{sec:heuristic}
In light of the patterns of resonant and near-resonant growth depicted and discussed in the previous sections, we expect changes in wave action that are no longer narrowly centered on resonances.  Ultimately, we wish to show how these changes signify more widespread changes in angular momentum taking place away from resonance.  To that end, in this section, we first devise a model spiral potential perturbation that we can eventually insert into a calculation of the torques exerted by short-lived long-wave spirals.  

The spiral we choose resembles most other, canonical spiral perturbations used extensively thoughout the literature, but now we introduce a time-dependent perturbation amplitude $F(R,t)$ that can vary with galactocentric radius $R$ and reflects growth and/or decay at rate $\gamma(R)$.  This growth and decay is assumed to occur on a timescale that is comparable to the dynamical time, although we comment on deviations from this scenario later.  For illustration purposes, we envision the radial variation of $F$ as a Gaussian centered at some radius $R_g$ with width $\Delta R_g$.  

Most generically, we write our spiral with radial wavenumber $k$ 
as a perturbation of the form 
\begin{equation}
\Phi_{sp}(k,R,z,t)=\sum_{m} \Phi_1(R,z,k,m,t)
\end{equation}
consisting of a number of $m$ components, although we will tend to assume that only one of these dominates in what follows.  

Following \citet{meidt22} we envision the perturbation as having some vertical extent tied to the disk's own vertical distribution (rather than being restricted to a 2D plane), i.e. 
\begin{equation}
\Phi_a(z)=\Phi_a(0) e^{-z/h_p}\cos{(k_z z)}
\end{equation}
where $h_p$ is comparable to the disk's own scale height.  We look in the smooth regime $k_z h_p\ll 1$ \citep{goldreich65}, which is the most conducive to growth localized to the mid-plane in the 3D disk, i.e. above the vertical Jeans length \citep{meidt22}.  Thus we have
\begin{equation}
\int_{-\infty}^{\infty}\Phi_a(z)dz\approx 2 h_p\Phi_a(0)
\end{equation}
An alternative would be to envision the perturbation as restricted to the plane with the form 
\begin{equation}
\Phi_a(z)=\Phi_a(0) e^{-\vert k \vert z}
\end{equation}
in order that it satisfies Laplace's equation away from the plane, as adopted by \citetalias{binney08} and \citet{goldreich78}. 
In this case, 
\begin{equation}
\int_{-\infty}^{\infty}\Phi_a(z)dz\approx \frac{2}{k}\Phi_a(0).
\end{equation}
In practice, a similar picture of growth in the plane is provided by either vertical form.  Our 3D perturbations makes it possible to account for the stability (and growth or decay) of the perturbation in the plane in relation to its vertical distribution, but in practice we let $k h_p$ be a factor of order of unity, 
with the expectation that this would be characteristic of the fastest growing 3D perturbations \citep{meidt22}.  

For any given $m$ the form of the perturbation in the plane is 
\begin{equation}
\Phi_1= \Phi_a(0) F(m,R,t) \cos{(kR+m\theta-\omega t)}.\label{eq:perturbation}
\end{equation}
Each $m$ has its own $F(m,R,t)$ describing the amplitude variation at that $m$ (discussed more below).

Explicitly including the factor $F(t)$ in the perturbation in deriving the torque below lets the $C_g$ calculation perform like a \citetalias{lynden-bell72} torque calculation: the growth and decay that we paste on with the function $F(t)$ depicts the change to the perturbation described in the most general \citetalias{lynden-bell72} calculation to second order in the perturbation (before the adiabatic limit $\gamma\rightarrow 0$ is invoked).  That is, in our second order torque calculation, which extends to first order in the perturbation, we include a time dependence that is the result of the second order changes examined by \citetalias{lynden-bell72}.  In this way, we can model where the growth and decay occur in the disk, as we leverage later in Section \ref{sec:diskchange}.  For example, a narrow $F$ centered tightly on corotation can be used to represent the torque in the adiabatic limit.  Meanwhile a broader $F$ profile provides a view of non-resonant torques in the non-adiabatic limit, without the need of performing a full non-adiabatic calculation in the manner introduced by \citetalias{lynden-bell72}. 

\subsection{A calculation of torques exerted by Transient Spirals}\label{sec:torques}
In this section we examine the impact of transience on angular momentum transport within the disk  
by calculating the torque
\begin{equation}
\tau=\int(\rho_0+\rho_{\rm sp})\frac{d(\Phi_0+\Phi_{\rm sp})}{d\phi}rdr d\phi dz
\end{equation}
exerted on the disk by the spiral outside
(as given in Section 6.1.5 of \citetalias{binney08}).  For our axisymmetric disk plus small first-order perturbation, this torque contains contributions from terms only starting at second order, i.e. 
\begin{equation}
C_g=\frac{R}{4G}\int_{-\infty}^{\infty}\int_0^{2\pi} \frac{\partial \Phi_{sp}}{dR}\frac{\partial \Phi_{sp}}{d\theta}d\theta dz.
\label{eq:Cg1}
\end{equation}
and only the terms in $\Phi_{sp}$ with the same $m$ make a non-zero contribution.  

Substituting the perturbation with form given by Eq.~\ref{eq:perturbation} into the expression for the torque in  Eq.~\ref{eq:Cg1} yields
\begin{equation}
C_g=\sum_m\frac{R}{4G} m (k h_p)\Phi_a(0)^2 F(m,R,t)^2 .
\label{eq:Cg2}
\end{equation}
(Note that a term proportional to $\gamma$ in the radial derivative of $\Phi_1$ integrates away.)  
The only major difference compared to \citetalias{binney08} is the explicit inclusion of $F(m,R,t)$.  
The factor $k h_p$ in parenthesis is absent when the density perturbation is 2D and restricted to the plane and will be neglected.

\subsubsection{Average torques and average changes in angular momentum}
In practice, the function $F(r,m,t)$ encodes the factors that distinguish the torques in Eq.~\ref{eq:Cg2} from the torques that lead to Eq.~\ref{eq:BT08Cg1} in the conventional steady spiral picture.  Now, when we write the total change in angular momentum due to gravitational torques over some time $t$ as   
\begin{equation}
L=\int C_g dt
\end{equation}
it will be limited by $F(t)$ to the spiral's lifetime.  

Consider, for example, a Gaussian time dependence \footnote{A second simple model might allow that the growth and decay occurs at two different rates, i.e. 
\begin{equation}
F(t)=\begin{cases}
e^{\gamma_g t} & \text{$t<t_s$}\\
e^{-\gamma_d t} & \text{$t>t_s$}\\
\end{cases}  
\end{equation}}
  
\begin{equation}
F(t)=e^{-(t-t_s)^2\gamma^2/2}
\end{equation}
consisting of growth and decay peaking at time $t_s$ over a timescale $1/\gamma$.   
The total change $\Delta L$ after time $t$, long after the peak in growth, is approximately $C_g \sqrt{\pi/2}/\gamma=C_g\tau$, in contrast to the change $C_g t$ brought about by a steady spiral.  

As a compact measure of the angular momentum exchanged by a transient spiral over the course of a dynamical time, or some generic period $2\pi/\omega$, let us now look at the average torque centered on the peak in spiral amplitude, 
\begin{equation}
\langle C_g\rangle =\frac{\omega}{2\pi} \int_{t_s-\pi/\omega}^{t_s+\pi/\omega} C_g dt
\end{equation}
In the case of the Gaussian time dependence, this yields
\begin{equation}
\langle C_g\rangle = \frac{R}{4 \pi G} m \Phi_{1,\rm peak}^2 \left[\frac{\sqrt{\pi}}{2}\frac{\omega}{\gamma } \textrm{erf}(\gamma\pi/w) \right]
\label{eq:Cgavg}
\end{equation}
where now for clarity we write $\Phi_a(0) F(t=t_s)$ as $\Phi_{1,\rm peak}$ and, for compactness, we replace the sum over $m$ by the single, dominant $m$ term associated with the number of prominent arms in our hypothetical spiral pattern.  

This expression returns the \citetalias{binney08} value in the limit of steady spirals, in which case the term in square brackets becomes 
$\pi (1-(\pi\gamma/\omega)^2)$ to lowest order in $\gamma/\omega$.  In the reverse limit, the torque falls off  proportional to $\omega/\gamma$.  Thus the shorter the spiral's lifetime $\tau=1/\gamma$, the smaller the average torque.  

At this point, our conclusion based on Eq.~\ref{eq:Cgavg} is that, as anticipated, transient spirals produce less change in angular momentum than long-lived spirals.  However, as discussed further below, it also encodes an important but less obvious change to where this angular momentum can be deposited: a series of transient spirals is capable of more widespread change in the disk angular momentum and structure than produced by steady spirals at narrow resonant zones.   

 \subsubsection{The Angular Momentum Current Carried by Transient Spirals}
 At first glance, Eq.~\ref{eq:Cgavg} might lead to the expectation that torques at resonance, where the wave grows and eventually decays \citep[][and \citetalias{meidt24}]{hamilton24}, are smaller than the torques away from resonance where the spiral is steady ($\gamma\rightarrow 0$).  However, this overlooks that the spiral's amplitude undergoes a larger change at corotation than it does away from it over a fixed period of time.  
 Centering the time average at all locations (radii) around $t_s(R=R_{CR})=t_{CR}$ instead of the time when the amplitude peaks at that radius we would have
\begin{eqnarray}
\langle C_g\rangle &=& \frac{R}{4 \pi G} m \Phi_{1,\rm peak}^2 \Big[\frac{\sqrt{\pi}}{4}\frac{w}{\gamma } \big(\textrm{erf}(\gamma\pi/w+\gamma t_s-\gamma t_{CR})\nonumber\\
&+&\textrm{erf}(\gamma\pi/w-\gamma t_s+\gamma t_{CR}) \big)\Big]
\end{eqnarray}
which can be written to lowest order in $t_{CR}/t_s$ and in the limit $\gamma\ll \omega$ as 
\begin{equation}
\langle C_g\rangle = \frac{R}{4 \pi G} m \Phi_{1,\rm peak}^2 \left(\pi-\pi t_s^2\gamma^2\right).
\end{equation}
The further from corotation, the further in the future $t_s$ is, and thus the smaller the average $C_g$ over the time surrounding the amplitude peak at corotation.    

A more succinct expression that captures differences in amplitude along a spiral would be
\begin{equation}
\langle C_g\rangle \approx \frac{R}{4 G} m \langle \Phi_1\rangle^2 
\label{eq:Cgavg2}
\end{equation}
where  the time-averaged amplitude over some averaging period $2\pi/\omega$ is
\begin{equation}
\langle \Phi_1\rangle\approx \Phi_{1,\rm peak}(1-t_s^2\gamma^2/2).  
\end{equation}

Although in this form, our expression for the angular momentum current in Eq.~\ref{eq:Cgavg2} appears almost identical to \citetalias{binney08}, there is an important difference:  whereas the latter implicitly assumes that the potential and density are steady and the same at all times, we explicitly allow that today's density and potential may be different than they were in the past (or will be in the future) and that the growth and decay may vary along the spiral.  In other words, the (potentially) differential growth and decay produces a spiral with its specific $\langle \Phi_1\rangle$, and any radial variations in $\langle \Phi_1\rangle$ formally translate via Eq.~\ref{eq:Cgavg2} into meaningful variations in $C_g$, as we consider in Section \ref{sec:appl}.  

From this point, assuming that the observed density (and by inference the potential) of any prominent spiral today must be representative of the average density and potential $\langle \Phi_1\rangle$, we can easily estimate the 
fractional spiral-driven angular momentum change after a period $\Delta t=2\pi/\omega$ as 
\begin{eqnarray}
\frac{\int C_{\rm g}dt}{L=2\pi \int \Sigma_0 V_c R^2}&\approx&
\langle C_g\rangle \frac{2\pi}{\omega}\frac{2G}{V_c^3R^2}\\
&\approx&\textrm{sgn}(k)\frac{m V_c}{2 R}\frac{\langle \Phi_1\rangle^2}{V_c^4}\frac{2\pi}{\omega}\nonumber\\
&\approx& \textrm{sgn}(k)\frac{m}{2(kR)^2}\frac{\Sigma_a^2}{\Sigma_0^2}\frac{V_c}{R}\frac{2\pi}{\omega}\nonumber\\
\label{eq:BT08Cg}
\end{eqnarray} 
using, for illustration purposes, the WKB approximation in the last line to write the potential in terms of the density, as in eq. \ref{eq:BT08Cg}.  
After N such periods, this becomes
\begin{equation}
\frac{\Delta L}{L}\approx \textrm{sgn}(k)\frac{\pi \tan^2 i_p}{m}\frac{\Sigma_a^2}{\Sigma_0^2}\frac{2\pi}{\omega}N_{}.\label{eq:deltaloverl}
\end{equation}

Note that here, as in the remainder of this work where the focus is on the torques exerted by long-wave, open spirals (with $kR\sim 1$ and $k<k_J$), we take the gravitational torque in Eq.~\ref{eq:Cgavg2} as the dominant contribution to the total angular momentum current.  It is straightforward to show that the advective torque
\begin{equation}
C_A=R^2\Sigma\int_0^{2\pi}v_{r,1}v_{\theta,1}d\theta 
\end{equation}
is at most the value calculated in the WKB limit \citep[see][]{binney08}.  The addition of terms outside the tight-winding limit in the ``Bottom's Dream'' cubic equation only lower it from this level.  When those terms dominate, the advective torque even reduces further in the regions of greatest interest, becoming proportional to $(\omega_e-m\Omega)$.  We thus proceed taking only the gravitational torque as a satisfactory, but still necessarily upper, estimate of the total torque.

\subsection{Changes in Disk Structure via Radial Variations in Angular Momentum Current}
\label{sec:diskchange}
As discussed at the beginning of Section \ref{sec:torques}, with the transience characteristic of material patterns and rigid waves in the long-wave limit comes the possibility of variation in $C_{tot}$ (or wave action), which 
in turn implies that the spiral may unevenly transport angular momentum.  
The result is an uneven loss of angular momentum from the disk interior to radius $R$.  According to the angular momentum equation and mass conservation, this loss translates to some mass flux that changes the mass contained with radius $R$ as
\begin{equation}
\frac{\partial M}{\partial t}=\frac{\partial}{\partial R}\frac{C_g}{V_c}
\label{eq:dMdt}
\end{equation} 
now focusing on a scenario in which the disk is a sub-dominant mass component whereby the mass distribution and not $V_c$ evolves in time.  

Changes in $C_g$, tracking changes in the shape of the spiral potential profile that arise with differential growth, thus give rise to local losses in angular momentum that drive mass inward. This represents a significant departure from the scenario predicted for steady spirals.  In those spirals, wave growth at corotation deposits angular momentum into the stars there and gives the spiral negative angular momentum (\citetalias{lynden-bell72}). Conserved wave action lets the wave carry that negative angular momentum, with no losses, up until it is absorbed at the ILR (where it can no longer propagate) at which point it must deposit its negative angular momentum.  Stars at ILR are thus forced to lose the angular momentum gained by the stars at corotation.  In contrast to these changes in wave action located entirely at resonances, in non-steady spirals changes in wave action are spread out.  The stars are thus forced to start losing angular momentum in the disk well ahead of the ILR, prompting widespread mass gain at inner radii through the associated inward flow of mass.   

\begin{figure*}
    \centering
    \hspace*{-.2cm}
    \includegraphics[scale=0.11]{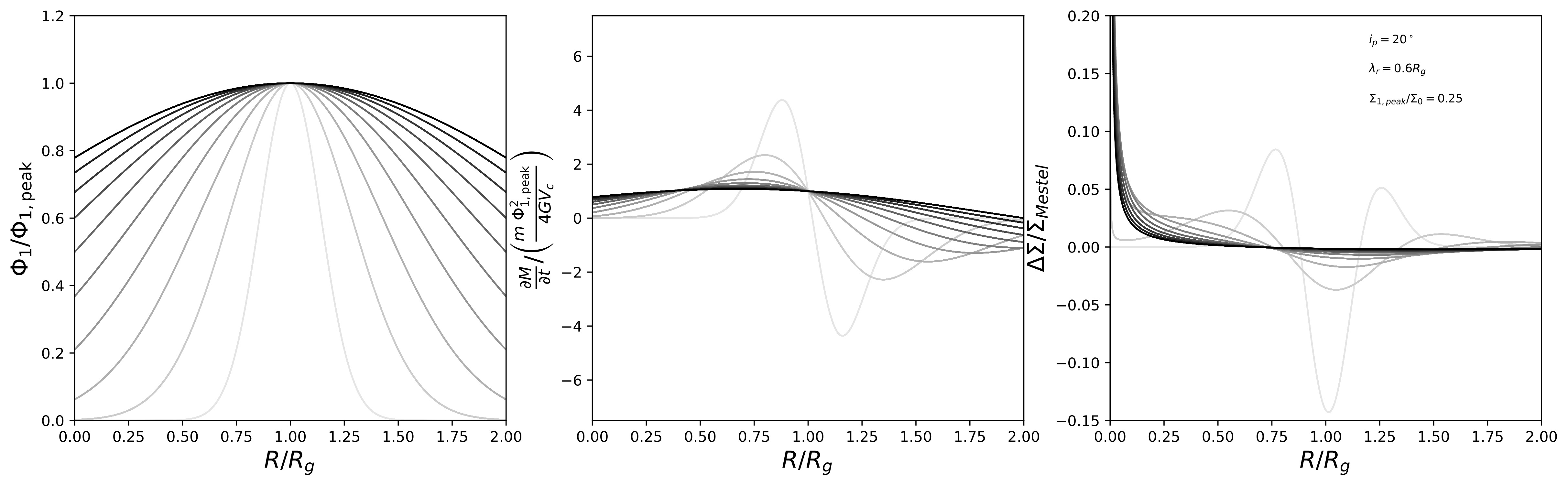}
    \caption{Illustration of the changes to radial profiles of disk mass (middle) and surface density (right) due to ten transient spiral waves (left) after one dynamical time. The spiral potentials in the leftmost panel are modeled as Gaussians centered on radius $R_g$ with width $\Delta R_g$.  The gray scale of each line varies with $\Delta R_g$, from low (light gray) to high (black).  The narrowest of these represents the growth in the adiabatic limit, while progressively larger $\Delta R_g$ illustrate the result of near-resonant, non-adiabatic growth.  Each potential is used to calculate $C_g$ according to Eq.~\ref{eq:BT08Cg}.  The mass and surface density evolution is calculated according to Eqs.~\ref{eq:dMdt} and \ref{eq:dSdt2}, respectively. The adopted spiral properties are shown in the legend in the rightmost panel.} 
    \label{fig:profiles}
\end{figure*}

\begin{figure}
    \centering
    \includegraphics[scale=.14]{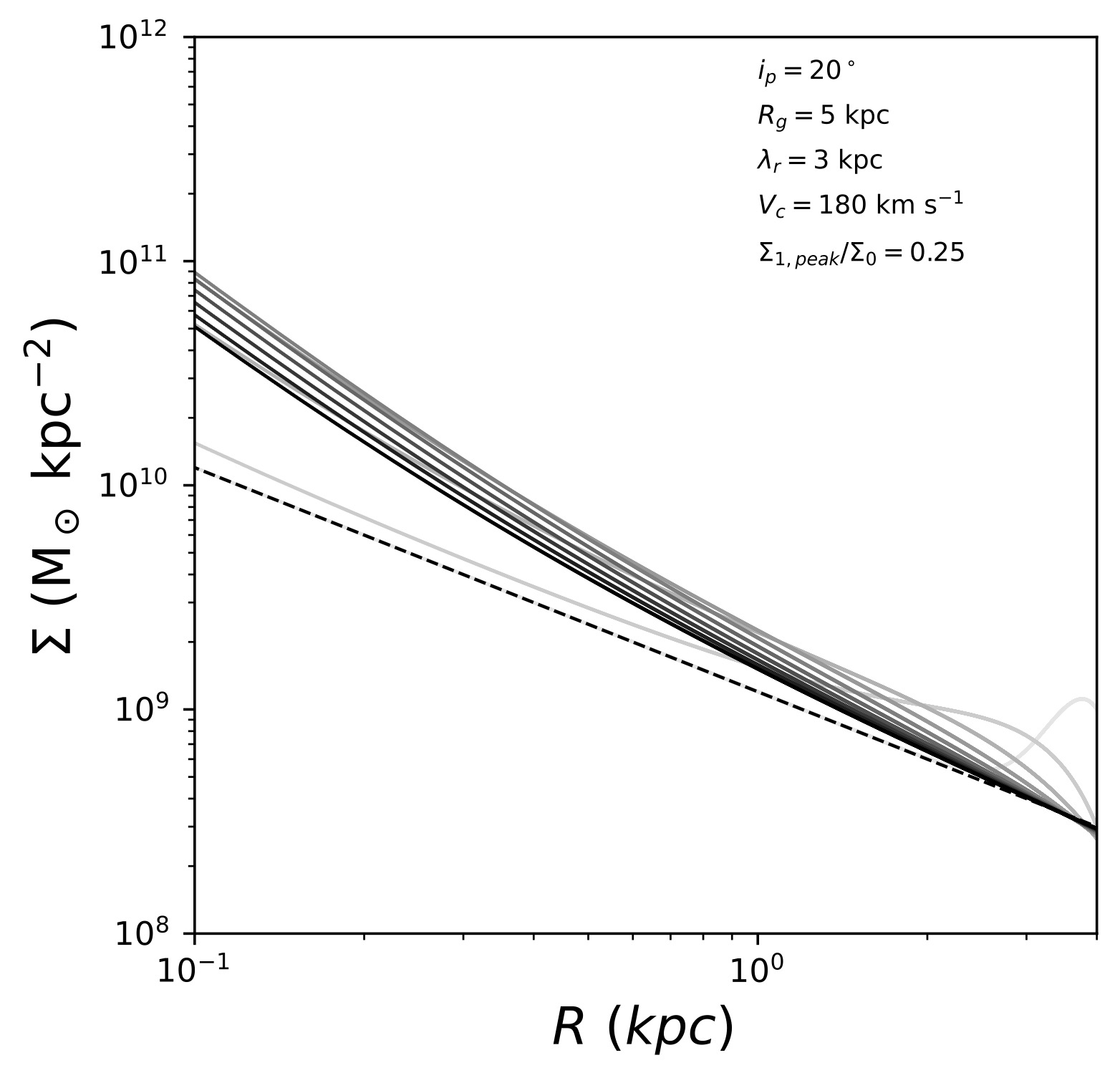}
    \caption{Inner surface density profiles (in physical units) predicted with the fractional changes in the right panel of Figure \ref{fig:profiles} after $N$=30 orbits have passed at $R_g$, adopting an idealized, Mestel disk configuration with a rotational velocity $V_c=180$ km s$^{-1}$ and setting $R_g=5$ kpc. The spiral is assumed to have a pitch angle $i_p=\SI{20}{\degree}$, and a radial wavelength $\lambda_r=2\pi/k=3$ kpc.  The total spiral duration is just over 5 Gyr. The grayscale of the lines is as in Figure \ref{fig:profiles}. The initial Mestel disk is shown as a dashed line. The adopted spiral properties are shown in the legend.}
       \label{fig:sigest}
\end{figure}

These changes can also be cast in terms of surface density evolution using that, after at time $\Delta t$, the change in surface density within radius $R$ is 
\begin{equation}
\Delta\Sigma=\frac{1}{2\pi R}\frac{\partial}{\partial R}\left(\frac{\partial}{\partial R}\frac{C_g}{V_c}\right)\Delta t.
\label{eq:dSdt1}
\end{equation} 

In this section we wish to get a sense of the magnitude and location of the changes that we can expect from observed spirals, given their typical pitch angles, arm multiplicities, and strengths. To this end, we begin by simplifying Eq.~\ref{eq:dSdt1} considerably to 
\begin{equation}
\frac{\Delta\Sigma}{\Sigma_{Mestel}}\approx\frac{V_c}{R}\frac{m}{4}\frac{\Phi_{1,peak}^2}{V_c^4}\frac{\partial}{\partial R}\left(\frac{\partial}{\partial R}(R\Phi_1^2/\Phi_{1,peak}^2)\right)\Delta t
\label{eq:dSdt2}
\end{equation} 
under the assumption that $V_c$ is roughly flat and using  $V_c^2=2\pi G\Sigma_{Mestel}R$.  Even though we do not expect Mestel disks to be conducive to spiral amplification (see previous section), below we will use the Mestel disk surface density as an effective surface density and plot the quantity $\Delta\Sigma/\Sigma_{mestel}$ to give sense of the magnitude of the surface density changes.  

It is also practical to translate the ratio $\Phi_{1,peak}^2/V_c^4$ into an expression for the spiral strength, given that only constraints on the spiral arm surface density amplitude (rather than potential) exist.  For this we rely on the WKB approximation, as in Section \ref{sec:torque2} and Section \ref{sec:shortwaves} and assume $\Phi_1\approx-4\pi G\rho_1/(k^2+m^2/R^2)$. This is not unreasonable for our present purposes, given that we are mostly interested in the peak magnitude of both quantities.  The $T_r$ and $T_{r,0}$ factors appearing in the full expression for the potential in Eq.~\ref{eq:poissonFull}, which we ignore in the WKB approximation, introduce a phase shift between the density and potential, and do not largely impact that peak magnitude.  Plus, adopting the WKB approximation here makes our results more easily comparable to other estimates.  

With these approximations, we estimate
\begin{equation}
\frac{\Delta\Sigma}{\Sigma_{Mestel}}\approx\frac{\tan{i_p}}{4(1+\tan{i_p}^2)}\left(\frac{\Sigma_{1,peak}}{\Sigma_{0}}\right)^2\frac{\lambda_r}{R}\frac{\partial}{\partial R}\left(\frac{\partial}{\partial R}(R\Phi_1^2/\Phi_{1,peak}^2)\right)\Delta N_{orb} R
\label{eq:dSdt2}
\end{equation} 
where we have written $\Delta t= N_{orb} 2\pi R/V_c$.  

Now we can inspect the changes brought about by any given potential profile $\Phi_1$ from Section \ref{sec:heuristic}, which is assumed to be the result of differential wave growth.  
The leftmost panel of Figure \ref{fig:profiles} shows plots of $\Phi_1/\Phi_{1,\rm peak}$ for ten such spiral arm profiles, each taking the form of a Gaussian centered on $R_g$ with width $\Delta R_g$.  These are meant to depict the product of different growth rates as a function of radius.  The middle panel shows the profiles of $\partial M/\partial t$ after one spiral lifetime, which we set to the orbital period at each radius (see Section \ref{sec:lifetimes}).   The rightmost panel illustrates changes to the disk surface density, plotted as the fractional change with respect to the effective disk surface density $\Sigma_{Mestel}$ after one dynamical time $R_g$ has elapsed.  As a reference, Figure \ref{fig:sigest} plots a zoom-in on the central densities calculated from the fractional changes shown on the right in Figure \ref{fig:profiles}, predicted in the case of a Mestel-like disk with a rotational velocity $V_c=180$ km s$^{-1}$ as observed in massive spiral galaxies. 

The lightest gray line adopts a small $R_g$ and depicts \citetalias{lynden-bell72} growth limited to resonance in the adiabatic limit of steadiness.  
Darker gray and black lines adopt progressively larger $\Delta R_g$ and illustrate the impact of non-steady open spirals that are able to grow to prominence away from corotation in a manner not captured by the adiabatic (steady, with $\gamma\rightarrow0$) limit adopted by \citetalias{lynden-bell72} but described in Section \ref{sec:longwaves}. 
The growth and decay in these cases is present over a larger radial range in the disk, allowing for prominent extended spiral arms.  
Whereas the small $R_g$ case illustrates the result of torques exerted by a single spiral, the more extended, prominent distributions with increasingly large $R_g$ more realistically portray scenarios achieved by the presence of multiple waves in serial, each with corotation at progressively lower $\Omega$ (at larger $R$) or a very extended shearing, material pattern.  (Arrangements like this are commonly found in simulations; e.g. \citealt{sellwood14,sellwood19}).  In this case, multiple generations of inner spirals could elapse during the lifetime of the outermost spiral.  For visualizing the change in surface density in either the right panel of Figures \ref{fig:profiles} or \ref{fig:sigest}, we therefore extend the calculation out to the dynamical time at $R_g$, rather than to one radially-varying dynamical time (single spiral lifetime).  

Although this set of models is certainly not an exhaustive representation of the spiral waves possible in disks, they illustrate several noteworthy behaviors.  First, 
in the small $R_g$ cases representing adiabatic, long-lived spirals, 
the changes predicted by Eq.~\ref{eq:dSdt2} perhaps not surprisingly lead to the creation of a density structure at corotation that resembles a 'groove'.  For these spirals, at locations away from corotation where $\gamma\rightarrow 0$, the torque calculation in Eq.~\ref{eq:Cgavg2} yields the picture of 'perfect' transport, and $dM/dt$ in Eq.~\ref{eq:dMdt} goes to zero.  

In the case of radially extended growth scenarios associated with a larger value for $\Delta R_g$, 
a wider change in $M$ and surface density is produced.  The narrow density feature that was created in the case of resonant adiabatic growth (modeled with small $\Delta R_g$) gets spread out as $\Delta R_g$ increases, resulting from mass flux as far in as the central-most radii.  The addition of even a little bit of mass at small $R$ can produce a large change in surface density, such that widely growing spirals very easily produce a compaction of the disk.    

The sort of angular momentum changes predicted here by assuming a larger $\Delta R_g$ are also apparent in the recent simulations of \citet{hamilton24c} in which a Mestel disk of stars responds to imposed spiral forcing that grows and decays in time along the full length of the spiral.  When the lifetime of the simulated spiral is as short as a dynamical time, the angular momentum changes produced around the spiral's peak are far wider than when the lifetime is longer.  The net azimuthally-averaged torque over the short lifetime also appears to fall slightly below zero in the region between the ILR and CR.  Spirals introduced into the simulations with long lifetimes, in contrast, exhibit net changes in angular momentum tightly arranged around the resonances.  In these simulations, the growth and decay are imposed, rather than produced as a result of the characteristics of the disk.

\subsection{Widespread dynamical heating}
\label{sec:heating}
Alongside the changes in the disk's mass and angular momentum distributions, spirals that grow to prominence also lead to dynamical heating of the disk (\citetalias{lynden-bell72}).  
For transient spirals, the energy and angular momentum densities of the wave vary in time, but the time-averaged energy and angular momentum, calculated by integrating over the disk from $R$ to infinity, are both still related as
\begin{equation}
E=\Omega_p J
\end{equation}
insuring conservation of Jacobi's integral.  As a result, just as for steady spirals, the change in energy for a given change in angular momentum is
 \begin{equation}
\Delta E=(\Omega_p-\Omega) \Delta J.
\end{equation}

 As described elsewhere, for steady spirals, the angular momentum gain for stars at corotation is not associated with heating, whereas the loss of angular momentum at the ILR leads to localized heating there.   
The changes in angular momentum at corotation due to transient ridigly rotating spirals also still yield no heating.  Now, though, the more widespread changes in angular momentum needed to support non-resonant growth spread the heating out.  

Widespread heating can also be produced by material patterns.  The continual evolution of the wave number in such cases implies that
\begin{equation}
E=(\Omega_p-2A) J
\end{equation}
and therefore that 
\begin{equation}
\Delta E=-2A \Delta J.
\end{equation}
Any loss of angular momentum brought about by a trailing material pattern, no matter where, heats the disk. 

Own its own, any single transient spiral  leads to only modest changes in $J$ and heating over the course of its lifetime.  But the combined impact of many such spirals present one after the other can be large.  

If one spiral lasts for 1-2 dynamical times (see Section \ref{sec:lifetimes}), then the net heating and angular momentum changes after N spirals can be comparable to the net heating produced by a single steady spiral after roughly N orbits.  Only these changes are more spreadout and not as tightly limited to resonances, as in the case of steady spirals.

\section{Observations: Spiral Fraction and Evolution with Redshift}\label{sec:emp}
 In this section we describe our empirical examination of spiral galaxies in the redshift range $z\approx 1-2.5$.  We will use constraints measured from the spirals to inform our modelling of the spiral-driven transformation process in Section \ref{sec:appl}.  
 
\subsection{Data and Sample Selection}\label{sec:emp_data}
The galaxy sample and its properties derive from the combined HST/ACS, HST/WFC3 and JWST/NIRCam deep field imaging mosaics across the wavelength range $0.8-4.5\mu$m produced by a number of surveys: CANDELS \citep{koekemoer11}, 3D-HST \citep{momcheva16}, CEERS \citep{finkelstein25}, JADES \citep{eisenstein23}, PRIMER \citep{dunlop21, donnan24} and COSMOS-Web \citep{casey23}. The raw data products were processed and analyzed by the Dawn JWST Archive (DJA)\footnote{\url{https://dawn-cph.github.io/dja/blog/2024/08/16/morphological-data/}} to provide the photometric redshifts, stellar masses, star-formation rates. We augmented these global properties with structural parameter measurements \citep{martorano25} and visual morphological classifications (this paper). For more details regarding the imaging datasets (fields, area, filter sets, depth) we refer to \citet{martorano25}.

For this paper we adopt the redshift range $1<z<2.5$ and a lower stellar mass limit $M_\star=10^{10.5}~M_\odot$. The redshift range ensures rest-frame near-infrared wavelength coverage for all galaxies. The mass limit ensures that high signal-to-noise ratio imaging with $S/N\gtrsim 200$ for all galaxies \citep{martorano24}, easily sufficient for the identification of spiral structure. For further analysis we only retain those galaxies with projected axis ratios $>0.35$ in the F444W filter (1841 out of 2380). For flatter/more edge-on galaxies spiral structure is difficult to ascertain.

\subsection{Classification Scheme}\label{sec:emp_class}

The visual classification is done in two steps. First, given limited person-power, one of us (AvdW) performed a quick-pass classification of the JPG cutout single-filter images and false-color images for all 1841 galaxies with the purpose to identify potential spiral galaxy candidates and reject the majority of galaxies that are obviously lacking in spiral structure, such as early-type galaxies with smooth light profiles and irregular morphologies such as mergers.  This produced 712 possible or likely spiral galaxies. 

For the formal classification both authors examined those 712 galaxies as well as a 200 randomly chosen other galaxies that did not qualify through the quick-pass test. The visually examined data products consist of JPEG images as shown in Figure \ref{fig:images}, and FITS images in a rest-frame optical filter and a rest-frame near-IR wavelength, allowing the classifier to change the contrast and brightness of the images with SAOImageDS9. The filters were chosen such that the same rest-frame wavelength is covered across the entire redshift range. Rest-frame $V$ band is provided by NIRCAM/F115W, F150W or F200W. Rest-frame $J$ band is provided by NIRCAM/F277W or F444W. The three filters to create the false-color images shown in Figure \ref{fig:images} are those that are closest to rest-frame $U$, $V$ and $J$. If an image is unavailable, the nearest medium or wide-filter image is chosen instead. 

The visual classification consists of answering the question: do you see spiral structure? The answer $S$ is a integer value ranging from 0 to 4 reflecting the classifier's confidence level that spiral structure is seen, with $S=0$ meaning 'certainly not' (corresponding to an indicative $<5\%$ confidence level evidence for spiral structure), $S=1$: 'probably not' $S=2$: 'maybe' ($40-60\%$ confidence),  $S=3$: 'probably', $S=4$: 'certainly' ($>95\%$ confidence). These percentages are, of course, indicative and subjective. If the answer to this first question is $S\geq 2$, then a follow-up question must be answered: how many arms are present? The answer, $A$, is an integer value of the number of visible arms. If the arms cannot be counted, for example because the contrast level is low, the arms are poorly defined, too short and many to be easily counted, then $A=0$.

With this scheme we identified 572 high-confidence spirals ($S\geq3$), and 161 possible spirals ($S=2$). For 683 out of these 733 (possible) spirals the number of arms could be counted: 82 1-armed spirals; 398 2-armed spirals; 123 3-armed spirals; 72 4-armed spirals, and 8 5-armed spirals. Examples of high-confidence ($S>=3$) spirals with counted arms ($A\geq 2$) are shown in Figure \ref{fig:images}. 

Out of the 200 randomly chosen galaxies that did not pass the first, quick-pass classification, 26 (13\%) have $S\geq3$ spiral structure. The reason for missing those initially is the examination of the FITS image with SAOImageDS9 as part of the formal classification scheme. Our calculated spiral fractions include an upward correction to account for this missing fraction, by assuming that 13\% of all unclassified galaxies have spiral structure. Any mass- or redshift-dependence in this factor is ignored.  It should be noted that no galaxies with effective radii smaller than $\approx 1$kpc were classified as spirals; this may be due to the limitation in spatial resolution.

The arm counts agree reasonably well with the recent results from \citet{espejo-salcedo25}, who also performed visual classifications based on the same NIRCam datasets as used in this paper. Our 2-armed spiral fraction (among all galaxies classified as a spiral galaxy) is 54\%, where they found 60\%. Our 3$+$ armed fraction is 28\% (compared with their 36\%). We classified 11\% as 1-armed spirals, a class that they did not define. One important difference between the samples is that our stellar mass limit is $M_\star=10^{10.5}~M_\odot$ and theirs is 0.5 dex lower.

\subsection{Spiral Fraction as a Function of Mass and Redshift}
\begin{figure*}
    \centering
    \includegraphics[scale=0.45]{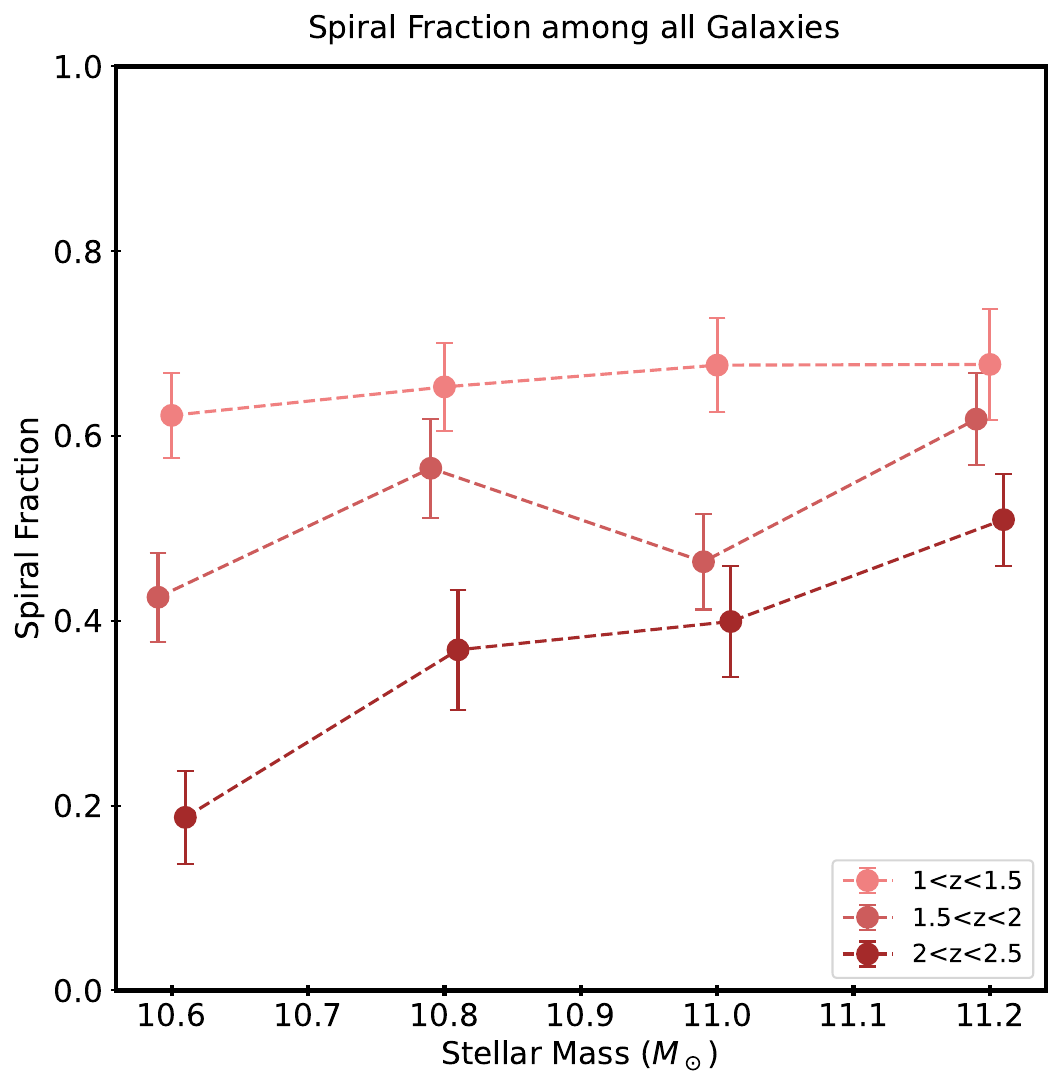}
    \includegraphics[scale=0.45]{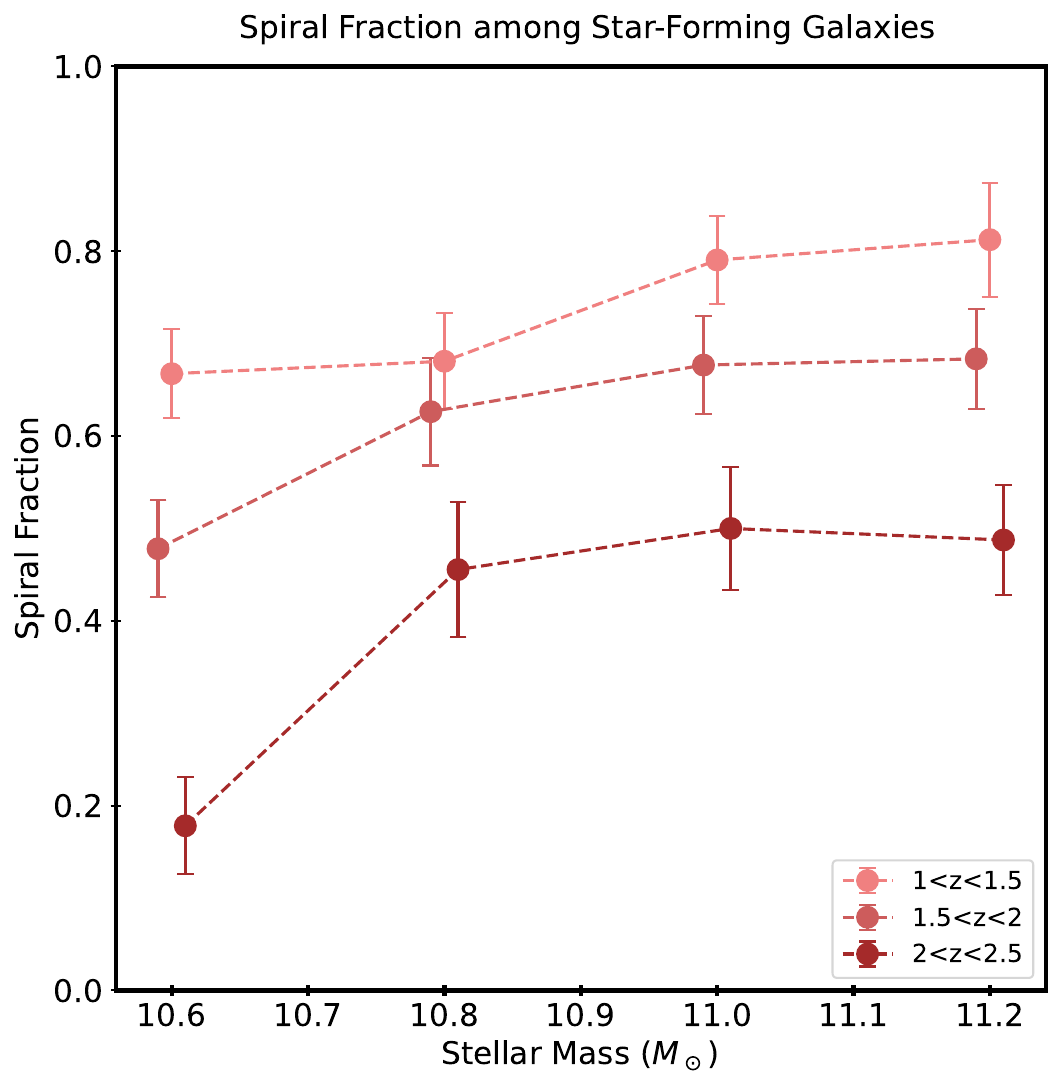}
    \caption{Left: Spiral fraction in the stellar-mass complete sample as a function of stellar mass and redshift. Right: Spiral fraction among star-forming galaxies in the the stellar-mass complete sample as a function of stellar mass and redshift.}
    \label{fig:spfrac}
\end{figure*}

The spiral fraction, its stellar mass dependence, and evolution with redshift are shown in Fig.~\ref{fig:spfrac}. The spiral fraction shows a weak dependence on galaxy mass, and a steady decline with redshift. Slightly more than 50\% of galaxies have spiral structure at $1<z<1.5$. 

Just like in the present-day Universe, spiral structure is far more common among star-forming galaxies. Considering just star-forming galaxies (right-hand panel of Fig.~\ref{fig:spfrac}) there is a more pronounced dependence on stellar mass. More than 80\% of $M_\star > 10^{11}~M)_\odot$~star-forming galaxies at $1<z<1.5$ have spiral arms, and even at $2<z<2.5$ this is still as high as $\approx 60\%$. Spiral structure is still common, if less so, at lower mass ($10.5 < \log{M_\star/M_\odot} < 11$), with $\approx 65\%$ at $1<z<1.5$ and $\approx 40\%$ at $2<z<2.5$. The mass dependence is likely related to disk settling: at increasing $z$, the number of settled disks decreases, especially at lower mass, as evidenced by kinematic measurements \citep[e.g.,][]{forster-schreiber06, kassin07, law07, wisnioski11, gnerucci11, newman13} and constraints on geometric shapes of galaxies \citep{van-der-wel14a, zhang19}. Despite the mass and redshift dependence of the fraction of galaxies with spiral arm structure, there is no evidence for a redshift or mass dependence of the arm multiplicity among spiral galaxies.

Our measured spiral fraction is somewhat higher than in \citet{espejo-salcedo25}. They find a spiral fraction (in the quiescent$+$star-forming galaxy population) of 55\% at $z\approx 1$ and 20\% at $z=2-2.5$. This can likely be explained by the difference in stellar mass limit.  Their sample includes galaxies down to $10^{10}~M_\odot$, where the spiral fraction is presumably lower given the trends in (Fig.~\ref{fig:spfrac}).

\begin{figure}
    \centering
    \includegraphics[scale=.45]{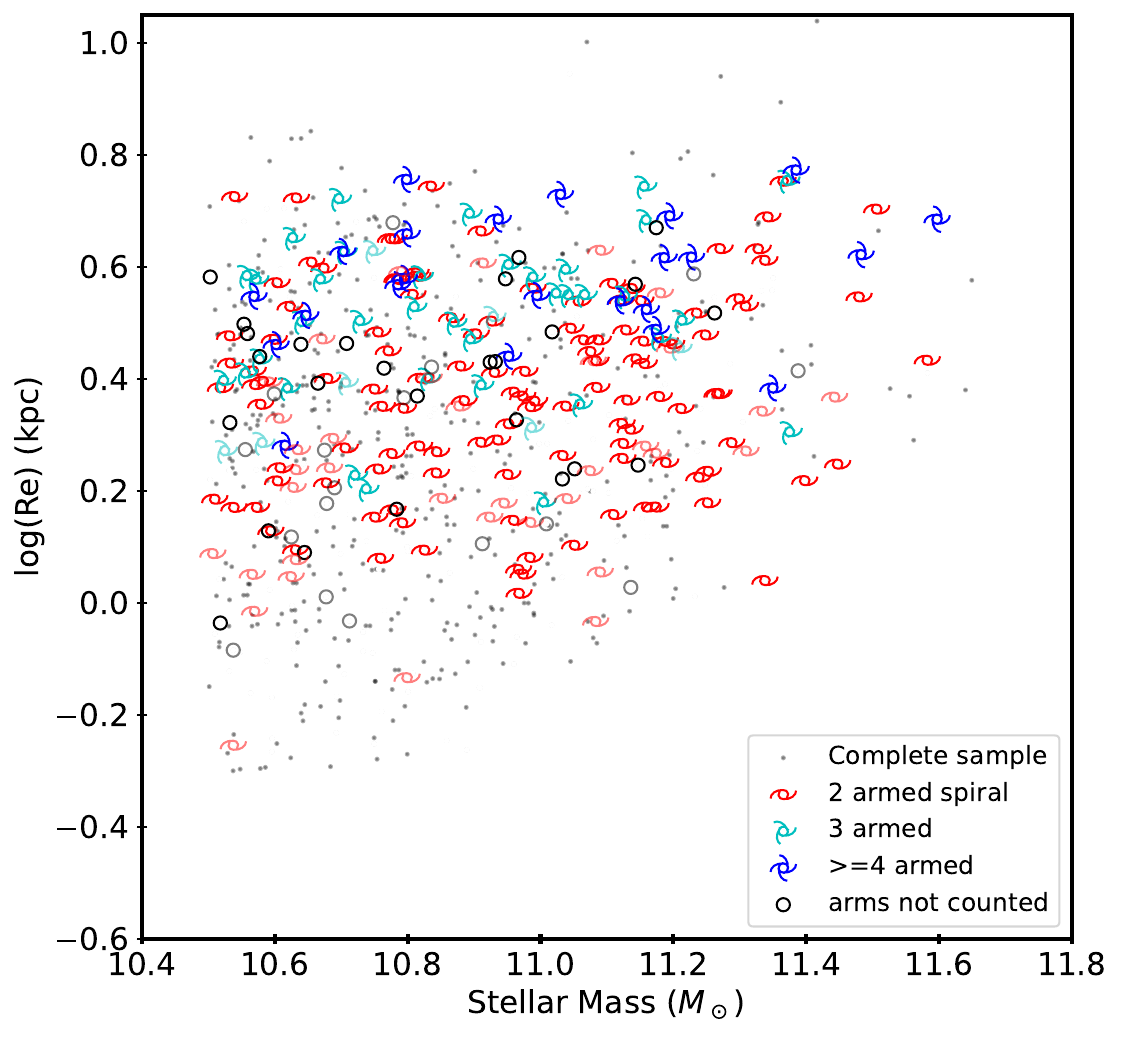}
    \caption{Galaxy size vs.~stellar mass for galaxies in the redshift range $1.5<z<2$, marking the spiral morphological type. Lighter-shaded symbols mark lower-confidence classifications (2).}
    \label{fig:arms}
\end{figure}

In Fig.~\ref{fig:arms} we show the size-mass distribution of the galaxies in our sample in the redshift range $1.5<z<2$, using the effective radius estimates at rest-frame 1.5$\mu$m from \citet{martorano24}.  There is no significant correlation between galaxy mass and the number of spiral arms, but the number of spiral arms varies systematically with galaxy size, with larger galaxies showing more arms. The scatter in size for galaxies with a given number of arms is non-negligible (0.1-0.2 dex), and the groups of 2-, 3- and 4-armed spirals do overlap.  This overlap presumably reflects true differences in the spiral structure among galaxies with the same effective radius together with some level of ambiguity in counting the number of arms. But, notably, the average sizes among the three groups are significantly different: for two-armed spirals the average effective radius is $2.4\pm 0.2$~kpc, while for three-armed and more-armed galaxies this is, respectively, $3.2\pm0.2$~kpc and $3.7\pm0.2$~kpc.  This simple observation implies immediately that spiral features reflect the properties of the underlying disk.

\begin{figure}
    \centering
    \includegraphics[scale=.5]{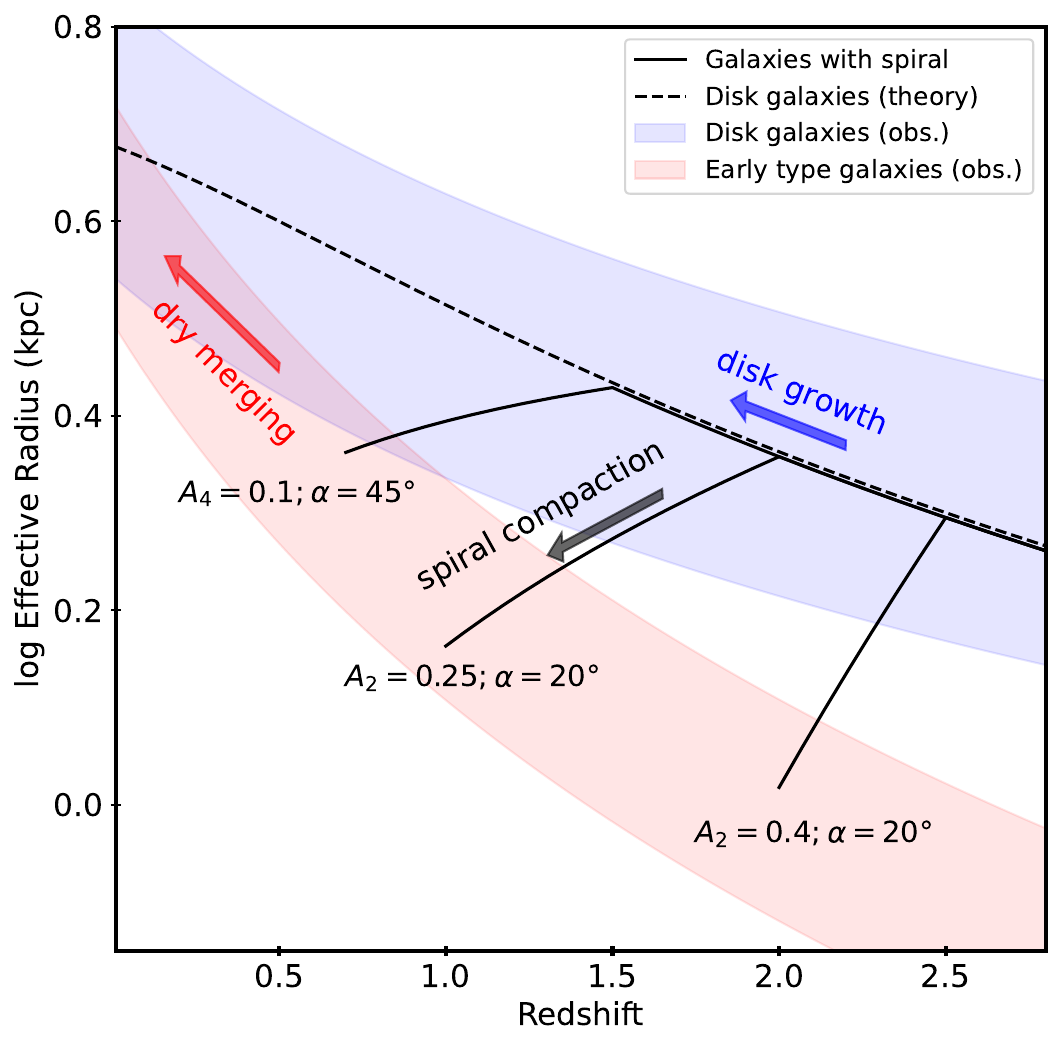}
    \caption{Galaxy size evolution tracks. In blue and red, the observed redshift dependence of size distribution of galaxies in the stellar mass range $10.5<\log(M_\star / M_\odot)<11$, separating between star-forming galaxies (our proxy for disk / late-type galaxies) and quiescent (early-type) galaxies, respectively. The fiducial, theoretical disk size evolution (dashed line) agrees well with the observed evolution of late-type/star-forming galaxies. To illustrate the feasibility of spirals to significantly alter the size evolution of individual galaxies, and transition from the the population of late-type to early-type galaxies, we show three examples of size evolutionary tracks due to angular momentum transport by spirals, with a variety of spiral arm contrast $A$, multiplicity $m=2-4$, and pitch angle $i_p$, as labeled. The structural transition (in size) is easily accomplished within a Hubble time.}
    \label{fig:Revol}
\end{figure}

\begin{figure*}
    \centering
    \includegraphics[scale=.45]{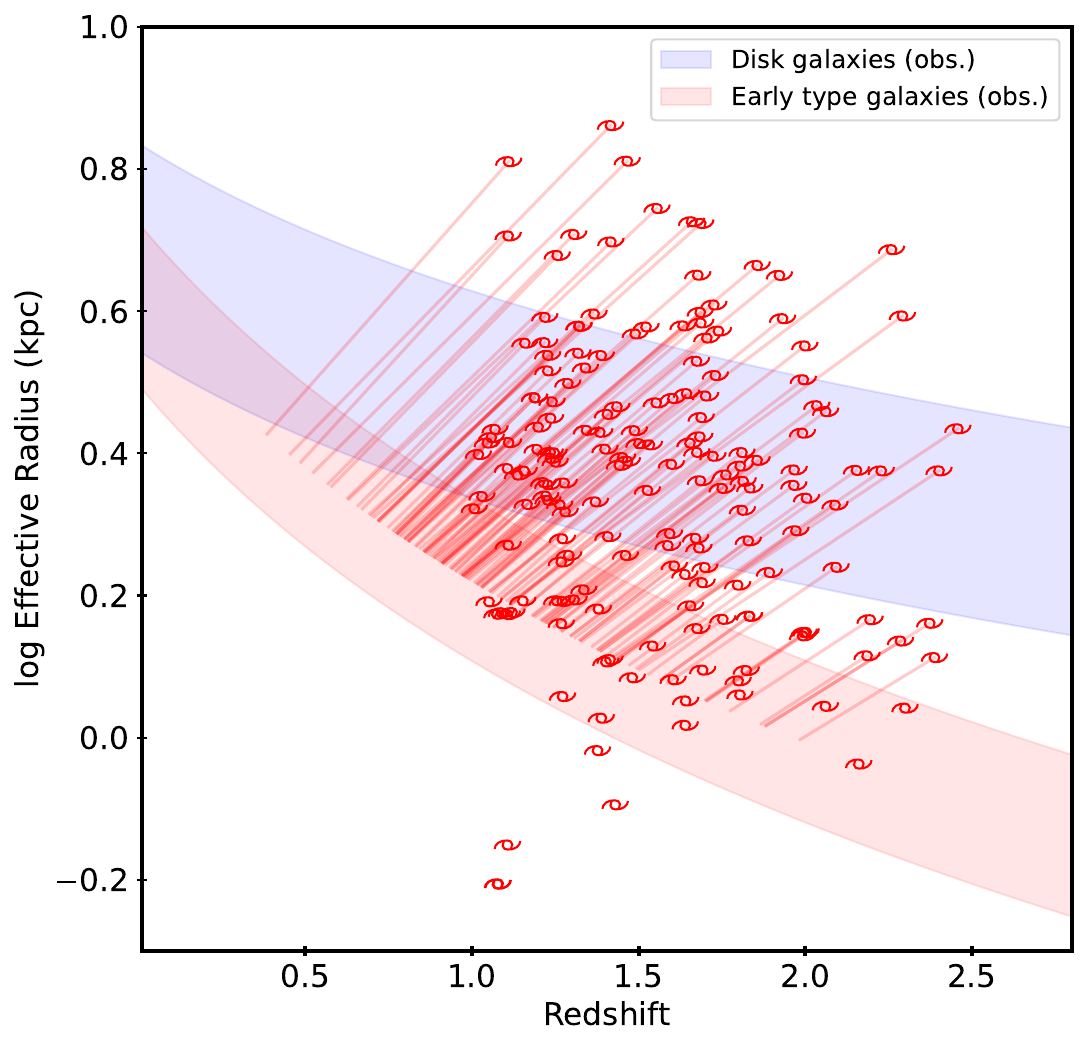}
    \includegraphics[scale=.45]{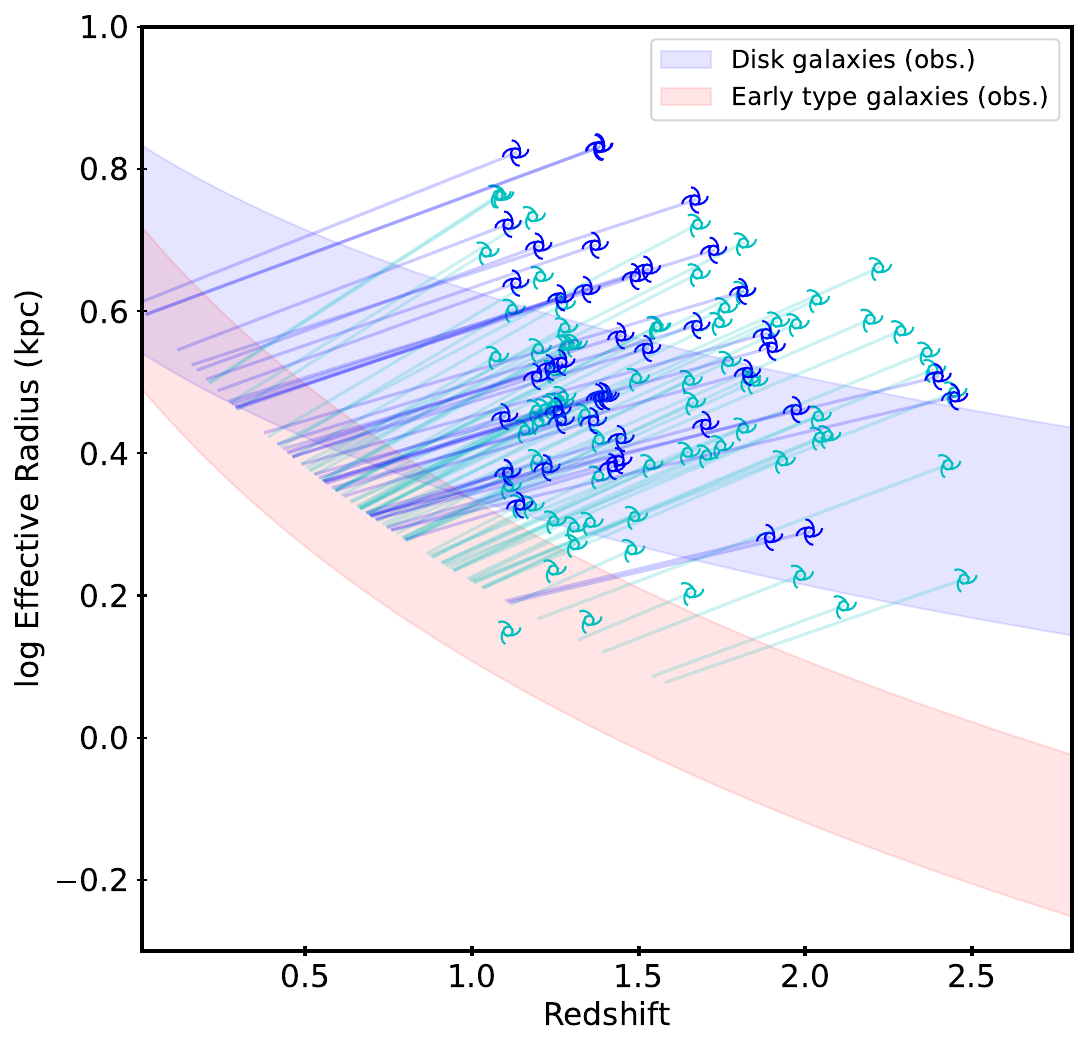}
    \caption{Similar to Fig \ref{fig:Revol}, we show the predicted size evolution of spiral galaxies (2-armed on the left; 3- and 4-armed on the right). The symbols represent the observed location in size-redshift, and the lines point to the evolutionary endpoint, taken to be centered on the size-redshift relation for quiescent / early-type galaxies.  In all cases the adopted amplitude and pitch angle are the same and unchanging ($A_2=0.3$ and $i_p=\SI{20}{\degree}$). Essentially all galaxies reach the size of an early-type-like galaxy size before $z=0$.}
    \label{fig:Revol_sample}
\end{figure*}

\begin{figure}
    \centering
    \includegraphics[scale=.45]{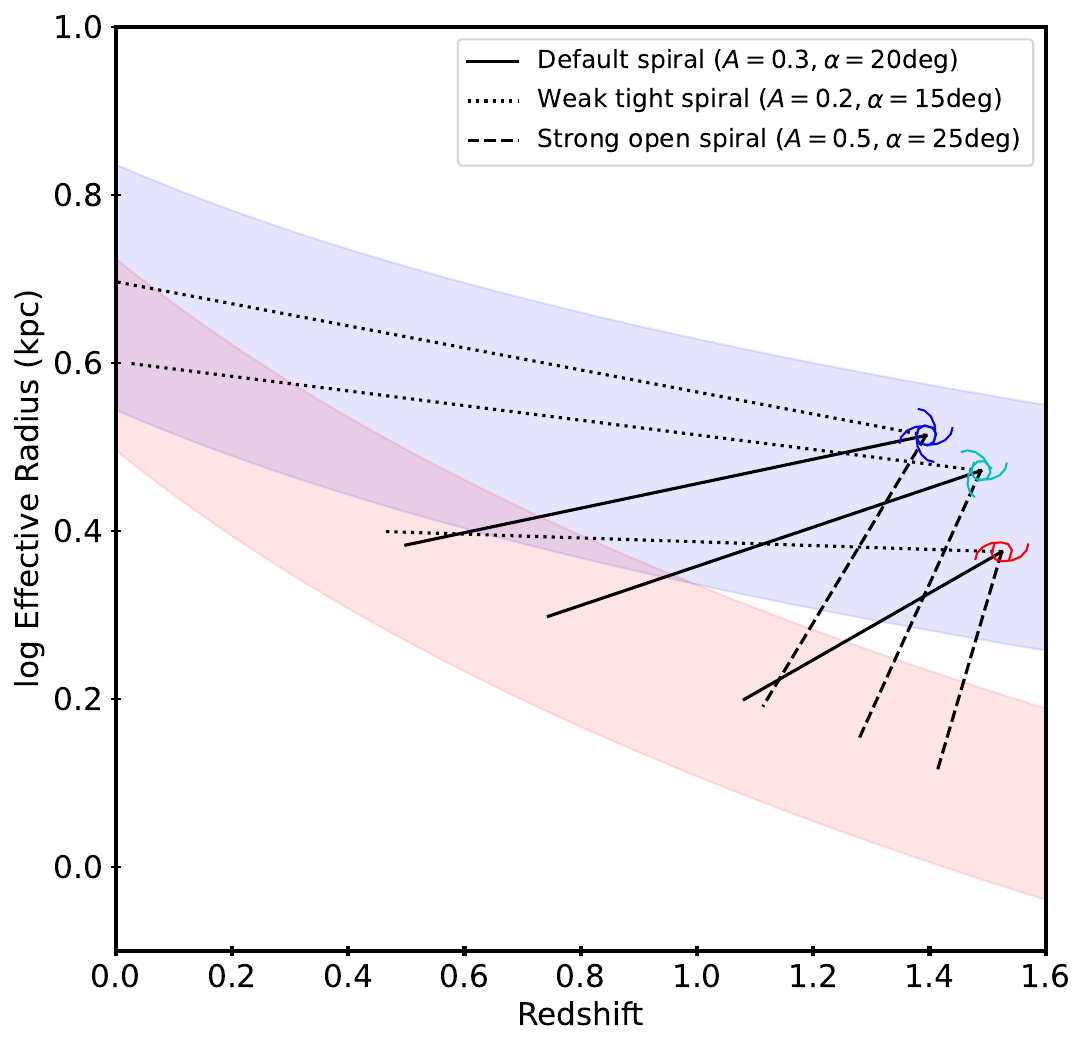}
    \caption{Similar to Figs.~\ref{fig:Revol} and \ref{fig:Revol_sample}, we show the predicted size evolution of spiral galaxies, now illustrating the sensitivity to spiral arm properties. Stronger spiral arms (dashed lines) lead to much larger, more rapid structural changes than weaker arms (dotted lines). The three spiral symbols at the start of each trajectory are placed at the median redshift and size of the three arm-number sub-samples in Figs. \ref{fig:Revol} and \ref{fig:Revol_sample}.}
    \label{fig:Revol_sens}
\end{figure}

\begin{figure*}
    \centering
    \includegraphics[scale=.45]{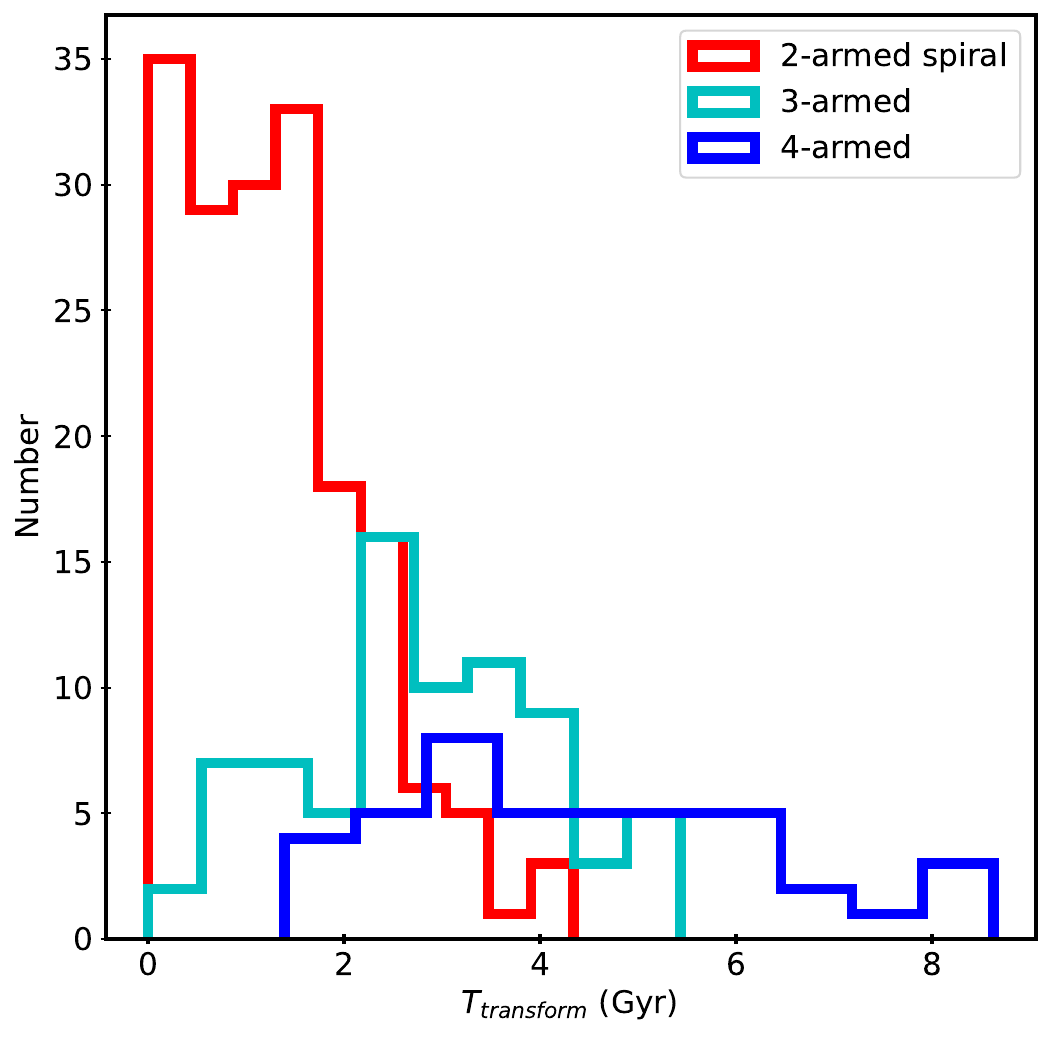}
    \includegraphics[scale=.45]{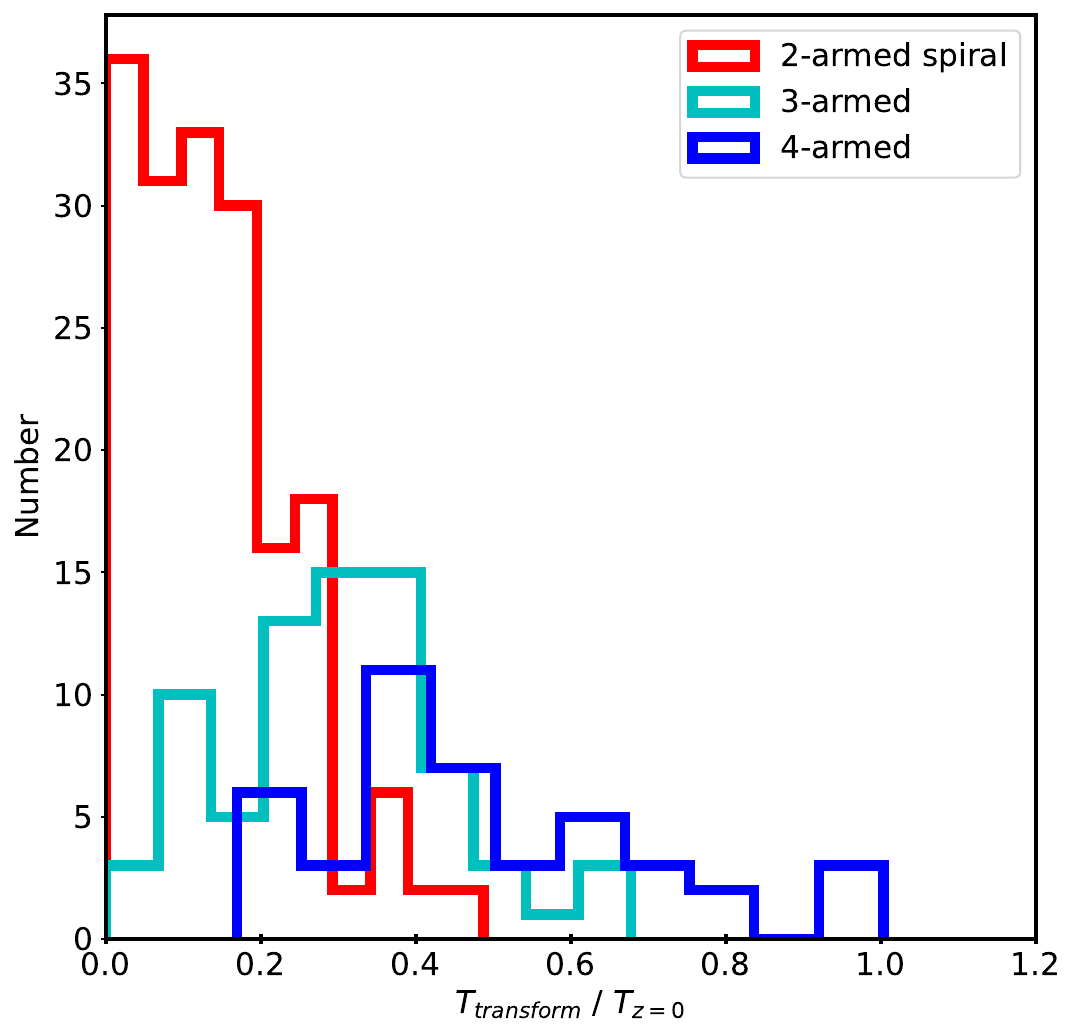}
    \caption{Distribution of transformation times, corresponding to the horizontal length of the lines in Fig.~\ref{fig:Revol_sample}. The transformation time is the time between the observed redshift and time the galaxy reaches the early-type galaxy size relation. Left: The absolute transformation time ranges from 0 Gyr (some spiral galaxies are already as small as the average early type) and 8 Gyr. Right: relative transformation time, compared to the time between the observed redshift and the present day. Essentially all values are $<1$, implying that all observed spiral galaxies at $z>1$ have sufficient time to transform into early-type galaxies by the present day, assuming that the spiral structure remains unchanged.}
    \label{fig:Ttr}
\end{figure*}

\begin{figure}
    \centering
    \includegraphics[scale=.5]{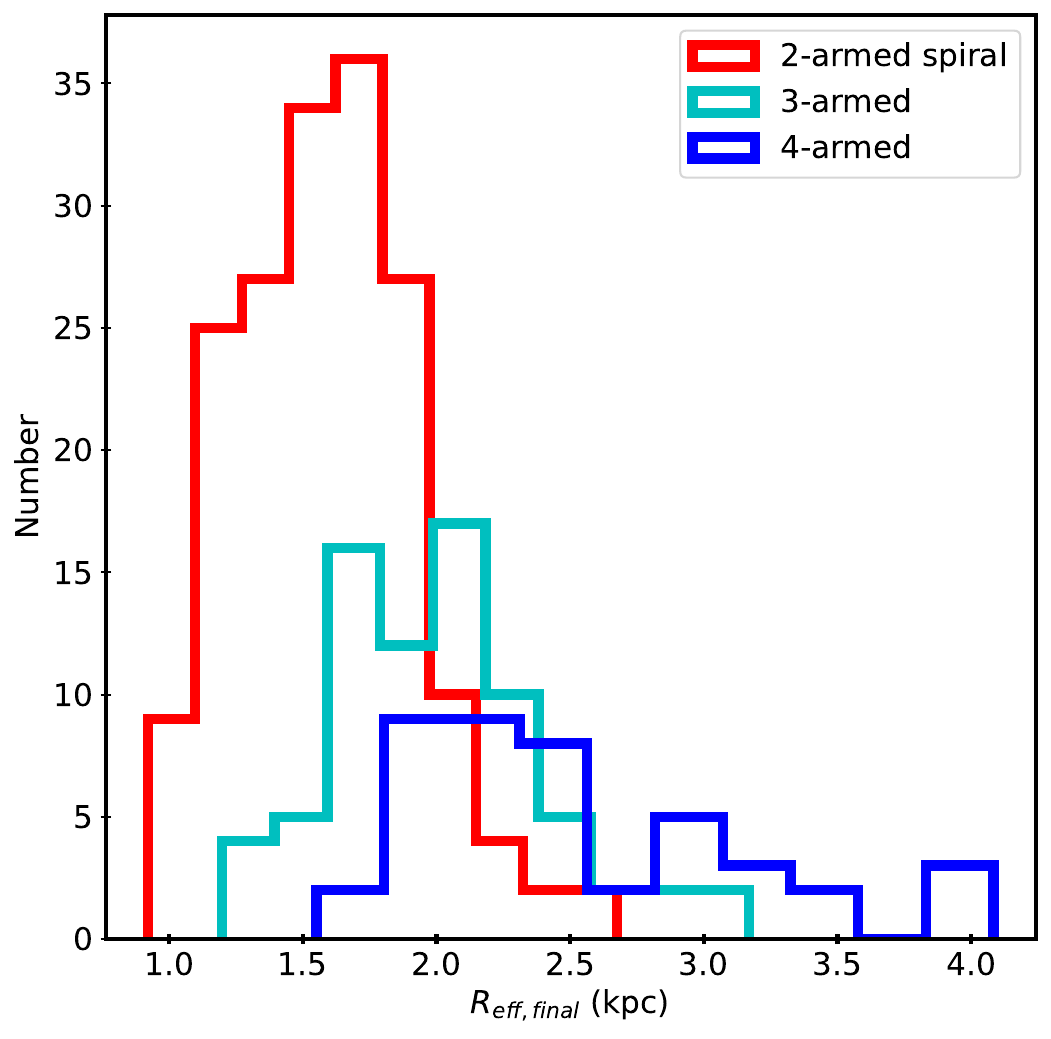}
    \caption{Size distribution of early-type galaxy descendants at the end of the transformation process (the end points of the lines in Fig.~\ref{fig:Revol_sample}.) Subsequent evolution (i.e., growth through merging) is not accounted for.}
    \label{fig:R}
\end{figure}

\section{The Impact of Spirals on Structural Evolution}\label{sec:appl}
\subsection{Size Evolution}
The spiral properties observed in the previous section provide the inputs needed to model spiral-driven transformation across cosmic time.  In what follows we will examine how galaxy sizes, in particular, are predicted to evolve.  
\subsubsection{Overview of the role of spirals}
We envision two avenues for angular momentum changes in disks: 1/short-lived non-steady patterns outside WKB approximation (whether a series of ridigly rotating modes or an extended material pattern) that drive broad changes in the mass and angular momentum distributions that build up through a succession of patterns over time and 2/classic long-lived steady patterns that induce changes at resonance but can lead to widespread changes in angular momentum through slow evolution in the pattern speed.  Note, though, that the calculations in Section \ref{sec:theory} suggests that transient waves, rather than steady waves, may be the more natural features to form in galactic disks.

As highlighted in section \ref{sec:torques}, the magnitude of the torque is the same in both of these two scenarios.  Besides differences in how wide angular momentum losses are achieved, though, there is a practical difference concerning the duration of each type of spiral.  In the first case, the integrated effect is determined by the number of orbits with the spiral present, while in the second case, the number of spirals and the duration of each is also relevant.  In section \ref{sec:lifetimes} we argued that each transient lasts for roughly a dynamical time.  We will thus make that assumption here, making the total number of orbits also the relevant measure of the integrated duration of recurrent, transient spiral waves.  

In Appendix \ref{sec:Norb} we show that the number of rotations for disks in a cosmological context can be estimated as

\begin{equation}
    N_{\mathrm{rot}}~(z_1<z<z_2) = 75~\ln{\left ( \frac{1+z_2}{1+z_1} \right)}.
    \label{eq:Nrotz2}
\end{equation}

Here, the orbital period is calculated for the effective radius, and should therefore be considered as reflecting the median orbital period for stars in a galaxy, assuming nearly circular orbits. The equation is derived in Appendix \ref{sec:Norb} for an Einstein-De Sitter cosmology and a correction for $\Lambda$CDM can be made by a simple fitting formula (Eq.~\ref{eq:Nadj}), but this correction is negligible for $z>1$. 

As outlined below, we use the $N$ in Eq.~\ref{eq:Nrotz2} to estimate the integrated angular momentum change.  During the full integration period the disk properties are assumed to implicitly vary in time as a result of spiral-driven structural changes.  In principle, the properties (pitch angle, number of arms, strength) should also vary in time, given their sensitivity to conditions in the underlying disk \citep[][\citetalias{meidt24}]{bertin89, bertin89a, toomre81}.  While it would be desirable to take that into account in a fully consistent model, such calculations are beyond the scope of the present work.  Here we aim instead for a basic illustration of the mechanism and keep the spiral properties fixed over the full period.

\subsubsection{Evolution in the effective radius}
In this section we describe how we apply our framework to model the observed size differences between late-type and early-type galaxies, ultimately developing a prediction for the change in the global size of galaxies (measured by the stellar effective radius) under the influence of spirals in a cosmological context. 

Following section \ref{sec:torques}, we envision changes in disk angular momentum prompting structural evolution through changes in surface density. This assumes that the evolving disk in question is embedded in a dominant halo, such that the galaxy rotational velocity stays constant over time. In this case 
we use Eq. \ref{eq:deltaloverl} to estimate the change in angular momentum.  
In practice, these make the additional assumption that the rotation curve is roughly flat.  

As discussed in Section \ref{sec:torques}, in deriving Eq. \ref{eq:deltaloverl} we need to assume a relation between the perturbed potential and density, given that empirical constraints on spiral arm strengths are in terms of the observable spiral arm density and its relation to the underlying (axisymmetric) disk surface density.  Several considerations suggest that it is reasonable to adopt the relation between density and potential in the WKB approximation (see Section \ref{sec:torques}).  
For more accurate modelling in the future, the choice of potential-density relationship should be kept in mind. 

Given our present goal of obtaining a compact, global measure of the spiral influence (as opposed to more detailed modeling of how the mass and angular momentum profiles vary within the individual galaxies) we select the disk size or effective radius as a proxy for surface density changes.  
There are several approaches to translate eq. \ref{eq:deltaloverl} into changes in effective radius with varying levels of sophistication.  
Given that our primary goal is a global accounting of angular momentum changes, we will take the simplest approach and assume that the rank order of stars (or gas) in terms of radius is unchanged: stars at the half-light light radius of a galaxy will remain at the (evolving) half-light radius. 
In this case, the time evolution of the half-light radius follows from the relative change in angular momentum at the (evolving) half-light radius. 
In this case, the relative decline in radius (i.e., angular momentum) for every rotation is
\begin{equation}
\delta_{\rm R}\equiv-\frac{dR}{R}=\frac{dL}{L} = \frac{\tan^2 i_p}{m} \pi A_{\rm m}^2 
\label{eq:dr}
\end{equation}

Consider a disk with an initial half-light radius $R_{\rm i} = 2$~ kpc, which is typical for $z=2$, and a trailing, two-armed spiral ($m=2$) with pitch angle $i_p=\SI{20}{\degree}$ and contrast $A_{\rm m=2} =0.3$. Every rotation, the relative decline in radius $\delta_{\rm R}\approx 0.02$.  
That is, after one orbit (at the half-light radius), the half-light radius is reduced by 2\% due to angular momentum transport.  

To translate Eq.~\ref{eq:dr} into an integrated measure of size change after $N$ orbits, there are again a few possibilities.  For example, when the goal is to inspect the result of a spiral with fixed properties on the initial size of the disk, then we would write 
\begin{equation}
R_{\rm f} = (1-\delta_{\rm R})N~R_{\rm i}.    
\end{equation}

Another approach, which we adopt below, is an attempt at a more conservative, realistic accounting.  For this scenario, we envision the disk and (unnormalized) spiral amplitude as adjusting after each rotation, supposing that the spiral $m$ and pitch angle remain otherwise unchanged for simplicity.  Thus, the aggregate effect of angular momemntum loss after $N$ rotations leads to a final radius
\begin{equation}
R_{\rm f} = (1-\delta_{\rm R})^N~R_{\rm i}.    \label{eq:sizeevol}
\end{equation}

Note that directly substituting the number of rotations at the effective radius from Eq.~\ref{eq:Nrotz2} into Eq.~\ref{eq:sizeevol} would ignore the reduction in orbital time due to the shrinking radius while keeping the orbital velocity constant. The actual number of rotations should increase with respect to Eq.~\ref{eq:Nrotz2}: like radius, the rotational period also decreases by $\delta_{\rm R}$.  That is, instead of a single rotation at the original radius $R_{\rm i}$ we have $1+\delta_{\rm R}$ rotations.  The result is 
a slightly more pronounced decrease in radius.  
Using that
\begin{equation}
(1-\delta_{\rm R})^{1+\delta_{\rm R}}\approx (1-2\delta_{\rm R}),
\end{equation}
in the limit of small $\delta_{\rm R}$, we write thus the reduction in radius as
\begin{equation}
R_{\rm f} = (1-2\delta_{\rm R})^{N_{\rm{rot}}}~R_{\rm i}.    
\label{eq:Rf}
\end{equation}
in terms of the original, unperturbed number of orbits $N_{\rm{rot}}$ from Eq.~\ref{eq:Nrotz}.   

Now letting transient but recurrent spiral structure persist from $z=2.5$ until $z=1.5$ (so that $N_{\rm{rot}}\approx 25$) in our example galaxy (with $m=2$, $i_p=\SI{20}{\degree}$, $A_{\rm m} = 0.3$, and $\delta_{\rm R}\approx 0.02$), Eq.~\ref{eq:Rf} implies that $R_{\rm f}/R_{\rm i} \approx 0.38 $, or that the radius decreases by nearly a factor 3 in this redshift interval.  Spirals can thus have a strong effect on structural evolution on a relatively short time scale.

\subsection{Size evolution in the cosmological context}
The last step needed to use Eq.~\ref{eq:Rf} to explain the evolution of galaxy sizes between the late-type and early-type segments of the population must account for the cosmological context of disk growth. If left unperturbed by spiral structure, disks evolve as $R\propto H^{-2/3} \equiv T_{\rm orb}^{2/3}=t^{2/3}$ at fixed mass (App.~\ref{sec:Norb}). Since $T_{\rm rot}=t/50$, the size growth in our fiducial disk evolution model is a factor $1+\delta_{\rm fid}=1.02^{2/3}\approx 1.0133$ per $T_{\rm rot}$. Our calculation needs to allow that the angular momentum and size increases that track disk growth happen continuously together with the spiral arm-driven transport of angular momentum outward.  Introducing $\delta_{\rm fid}$ into Eq.~\ref{eq:Rf}, we thus write the combined change in size as
\begin{equation}
R_{\rm f} = (1-2\delta_{\rm R})^{N_{\rm{rot}}}~R_{\rm fid} = (1-2\delta_{\rm R}+\delta_{\rm fid})^{N_{\rm{rot}}}~R_{\rm i}.
\label{eq:Rf2}
\end{equation}

Adopting the properties of our fiducial $m=2$ spiral (as in the example in the previous section), the combined effect of disk growth in a cosmological context plus compaction due to spiral-driven angular momentum transport suggests $R_{\rm f}/R_{\rm i} \approx 0.54$ from $z=2.5$ until $z=1.5$.  This is a slight but significant upward correction from 0.38.  

This final picture of evolution is illustrated in 
Figure \ref{fig:Revol}, where we show three examples of evolutionary tracks due to spiral compaction in the context of the observed size evolution of the galaxy population as a function of redshift. We note that the observed evolution of star-forming galaxies (taken as a proxy for late-type galaxies) from \citet{martorano24} agrees well with the fiducial disk growth model invoked in writing Eq.~\ref{eq:Rf2}. As illustrated by each of the three spiral compaction tracks (adopting variety of spiral parameters), spirals reduce  galaxy sizes enough to make them consistent with the observed sizes \citep{martorano24} of quiescent (non-star forming) galaxies, which we take as a proxy for the early-type population. 

We apply our spiral compaction calculation directly to observed spiral galaxies in Figure \ref{fig:Revol_sample}. The size distribution of the spiral galaxies in this observed sample is consistent with that of star-forming galaxies and is generally larger than the sizes of early-type galaxies.
For each track we adopt the observed number of spiral arms and use the measured size as the initial disk size.  Lacking measured spiral pitch angles and amplitudes for our sample, in Fig. \ref{fig:Revol_sample} we choose to adopt the same value for all and then later illustrate the sensitivity of the results to the chosen spiral properties in Fig. \ref{fig:Revol_sens}. Our default values for arm amplitude and pitch angle are: $A_m=0.3$ and $i_p=\SI{20}{\degree}$. The latter value is consistent with the average value of the $z>1$ spiral pattern fits from \citet{chugunov25} based on some of the same NIRCam data as used in this paper. Those authors do not directly estimate the amplitude, but rather the luminosity of the spiral component, and a conversion to $A_m$ is not possible. Instead we take the typical value of $A_m=0.3$ for local spiral galaxies from \citet{elmegreen11}. 

With these adopted properties, we let the duration of the spiral compaction phase (or the transformation time) for each individual track last until the size reaches the midpoint of the early-type galaxy size distribution (the end points of the lines in Fig.~\ref{fig:Revol_sample}).  
From the redshift location of the endpoint of each track we see that most 2-armed spirals reach the midpoint of the early-type galaxy size distribution before $z=1$ and essentially all galaxies do so before $z=0$.  The implication is that there is more than sufficient time between the redshift of observation and the present-day (i.e. a sufficient number of rotations) to achieve a structural transformation from late- to early-type.

The distribution of calculated transformation durations is shown in Figure \ref{fig:Ttr}. The duration times 
are taken from the redshift intervals of the lines in Fig.~\ref{fig:Revol_sample}. Typical transformation times are $1-2$~Gyr for the 2-armed spirals, up to $4-6$~Gyr for the 4-armed spirals and, as noted above, nearly always shorter than the time between the redshift of observation and the present day (see the right panel of Figure \ref{fig:Ttr}). By construction, the sizes of the galaxies at the end of the transformation agree with the observed sizes of early-type galaxies at those epoch.  We show these in Figure \ref{fig:R}, as an accessory to the histograms of transformation time.   

Let us now consider the sensitivity of these evolutionary tracks and time scales to variations in the spiral arm properties. Compared to the spiral properties adopted in Fig. \ref{fig:Revol_sample} (and Figs. \ref{fig:Ttr} and \ref{fig:R}), many present-day spirals are much more prominent with $A_m\approx0.5$, especially  the 2-armed grand-design spirals, while others are weaker with $A_m=0.2$. Spiral pitch angles are also observed across a range of values, typically $i_p=\SI{15}{\degree} - \SI{25}{\degree}$. This variety in the patterns is present in the galaxies in our $1<z<2.5$ sample (Fig.~\ref{fig:images}). The quadratic dependencies on $\tan{i_p}$ and $A_m$ in Eq.~\ref{eq:dr} imply a strong impact on the spiral torques and associated size changes, which we illustrate in Fig.~\ref{fig:Revol_sens}. The strongest spiral arms (dashed lines) can act much more rapidly than depicted in Fig.~\ref{fig:Ttr}, essentially instantaneously transforming a galaxy, akin to violent disk instabilities (see Discussion): as $A_m$ approaches unity, the time scale for transformation approaches the orbital time ($N_{\rm{rot}}\sim 1$). Even weaker spiral arms (dotted lines) can still (if secularly) lead to a transformation, but this happens on a Hubble time scale.  The weakest spirals will act primarily to slow down the net size growth of the disk, rather than leading to compaction.

It is worth emphasizing that size evolution depicted for individual galaxies in Fig.~\ref{fig:Revol_sample} should be interpreted within the context of rapidly evolving populations. The spiral compaction trajectories have galaxies `arrive' in the early-type galaxy population at an increasing rate from $z=2.5$ to $z=1$: the number of spiral galaxies can be used to predict the growth rate of the early-type galaxies. At the same time, new spiral galaxies will emerge in this mass range through star-formation. The declining arrival rate from $z=1$ to $z=0$ in Fig.~\ref{fig:Revol_sample} reflects this missing aspect. A more general description of structural changes in this context requires prescriptions for evolution in stellar mass and star-formation rate.  

Another point worth emphasizing is that our proposed evolutionary channel specifically models the production of disky (flattened), rotating early types not classical, triaxial ellipticals.  
The former class of objects (with low star-formation rates, centrally concentrated, and dynamically hot disks) vastly outnumber classical ellipticals in the present-day Universe \citep[e.g.,][]{emsellem07} as well as at earlier times \citep{chang13a, van-houdt21}. It is only among the rare very high mass ($M_{\star}>10^{11.3}~M_{\odot}$) that classical ellipticals dominate in number \citep{van-der-wel09a}. 
These likely form through dissipationless merging \citep[e.g.,][]{bezanson09, van-der-wel09, hopkins09a, newman12, bluck12, van-der-wel14}.  
That is, early-type galaxies are expected to continue their evolution after their transformation from star-forming spiral to disky early-type, but through the \textit{ex situ} growth rather than \textit{in situ} processes.

As an aside, internal torquing by spiral structure as an evolutionary process is not an alternative to the transformation of passive disks through environmental processes, which are limited to massive groups and clusters. The majority of early-type galaxies exist outside such dense environments \citep{van-den-bosch08b, van-der-wel10}, and internal processes must be invoked to explain their evolution \citep[e.g.,][]{baldry06, van-der-wel08, peng10a}.

\section{Discussion}\label{sec:discussion}
\subsection{Structural changes in galaxies}
In the previous section we showed that the emergence of strong, low-$m$ spiral structure in stellar disks predominantly at high $M_\star$ drives angular momentum outward and leads to sufficient compaction to reproduce the observed sizes of early-type galaxies. The effective radius served as an ideal compact measure of the transformation and made it possible to compare our predictions with observed galaxy sizes. But structural evolution is not simply a reduction in size; the entire density profile changes in shape. We envision this as taking the form  of central mass build-up (bulge formation) and a steepening of the profile toward higher S\'ersic index \citep[as observed, ][]{martorano25}. At the same time, the outward transportation of angular momentum also heats the disk (\citetalias{lynden-bell72}; Section \ref{sec:heating}), increasing the velocity dispersion. 

These are precisely the ingredients that are needed for a complete description of the transition from late- to early-type.  We envision a history in which disks form most of their stars during a main formation epoch, which comes to a close as stellar spiral arms activate, reducing the half-light radius and increasing the concentration (due to torques) while also increasing the thickness of the disk (due to heating).  This is consistent with the observations of $z\approx 1$ galaxies, where early-type galaxies have larger velocity dispersions than late-type galaxies \citep[e.g.,][]{van-der-wel21} but generally retain their disk-like structure as flattened, rotating stellar bodies \citep{van-houdt21}. 

Eventually, the very changes brought about by spirals inhibit their formation (section \ref{sec:fate}). Besides an angular momentum criterion, spirals will also cease to exist once the Jeans length increases beyond the size of the galaxy. Indeed, for early-type galaxies the Jeans length exceeds their effective radius: there is no "space" for spiral arms. This process can explain the large range in structural and morphological properties seen among $10.5 < \log{M_\star/M_\odot} < 11$ galaxies, describing an evolutionary sequence for thin disks with high-$m$ gas spirals, to stellar-dominated disks with low-$m$ stellar spirals, to early-type galaxies with smooth light profiles.

Structural transformation notably goes hand in hand with a decline in star-formation activity  \citep{kauffmann03a, franx08, bluck14, whitaker15, barro17, martorano25}.  According to a number of studies, we can trace the coincident evolution of structural change and `quenching' \citep[e.g.,][]{faber07} to the coincident growth of the central black hole \citep[e.g.,][]{chen20}. Here, too spiral arms may play a key role: torques exerted by strong, low-$m$ stellar spirals on the gas disk can drive large amounts of gas to the center, feeding the central black hole. The specifics of the mechanism that allow the black hole to quench star formation do not need to be discussed here, but it seems plausible that the black hole, once grown sufficiently massive, prevents or slows the cooling process in the halo, thereby cutting off the replenishment channel for the cold ISM.

The structural transformations that can arise with EASE have been previously attributed to a number of other mechanisms.  What sets EASE apart from other compaction scenarios is that it makes stellar spirals, rather than instabilities in the gas disk \citep{noguchi99, bournaud07a, elmegreen07, genzel08, dekel09}, a major avenue for reshaping disks. 

At the same time, the EASE mechanism is also expected to be  active in gas disks.  
The spiral features that form in cold(er) gas disks are predicted to be more filament like, with very high $m>30$, as confirmed by JWST/MIRI observations of nearby galaxies \citep{meidt23}.  According to eq. \ref{eq:BT08Cg}, these gas features  produce very little torque on the stellar disk, given that high-$m$ stellar perturbations remain negligible, and thus drive little angular momentum changes in the stellar disk.  However, their influence on gas disks is much greater. The torque increases with $m$, which will tend to offset the weakening of gas perturbations present in gas disks that represent a smaller mass component in the galaxy.  In gas-rich, higher-$z$ galaxies, the torque can be even more significant.  Thus, the same process of non-resonant growth depicted in this work would generate strong angular momentum and mass redistribution in gas disks, fueling the build-up of a central gas concentration and, through starburst activity, a central, compact stellar component. Indeed, dusty, star-bursting cores are seen in massive, star-forming galaxies at $z\approx 2$ \citep{tadaki15, le-bail24, gebek25}, but not at lower $z$.

In the next section we make a more detailed comparison between  
the gas flows predicted with EASE and the process of clump formation and clump migration thought to lead to compaction. 

 \subsection{Relation to other forms of compaction}
Our line of reasoning and calculation shares conceptual and physical similarities 
to the clump-driven compaction narrative of \citet{bournaud07a} and \citet{dekel09}: departures from axisymmetry form and exert torques that are sufficiently strong, and act sufficiently rapidly, to change the overall structure given the short orbital times of galaxies at $z\approx 2$ \citep[also see][]{genzel08}. 
In our picture (EASE) spiral arms with contrast of at least $A_2=0.2$ play that role,  
whereas in \citet{dekel09} it is satisfied if at least 20\% of the disk mass is in the form of clumps.

There are, however, several fundamental differences. In the original compaction picture, gas-rich, Toomre unstable disks produce massive, gravitationally decoupled clumps. The resulting torques make the clumps themselves migrate to the center.  In EASE, disks form transient spiral arms not through disk-wide Toomre instability, but instead through an easy, more gentle, almost continuous process that leverages self-gravity and differential rotation.  Those spiral arms exert torques that make disk material migrate to the center.  In EASE, both stellar and gaseous disks experience spiral arms and torques, whereas clumps emerge and viscous torques act only in gas disks.  
These differences are important. Clumps only form at early times, at $z\gtrsim 1.5$, when gas fractions were sufficiently large. Any structural evolution beyond $z\sim 1$ would need to be driven by a different process.  

In our picture, spiral arms represent an avenue for structural evolution across the whole redshift interval $0<z<2.5$, and even beyond, whenever a settled disk emerges.  What changes from $z\sim 2.5$ to $z=0$ is where the torques are located and which disk component primarily evolves: the low gas fraction at the present day implies that the stellar body contracts, whereas the high gas fraction at $z\approx 2$ implies that, primarily, gas will be transported toward the center, as observed by \citet{genzel23}, producing compact, star-forming (and dusty) centers \citep[as observed, e.g.,][]{tadaki15, miller22, gebek25}, and potentially feeding the black hole. But the mechanism is the same throughout cosmic time, even if at early times the process is accelerated due to shorter orbital times.

Future work, where direct measurements of torques due to spiral arms and clumps are compared, will determine whether gas$+$young star clumps or stellar spirals dominate the compaction process. The difficulty with this phenomenological approach is that it is not straightforward to empirically distinguish between clumps and spirals. In practice, similar techniques are used to identify both sorts of structures. \citet{kalita24} measure clump fractions of $1<z<2$ with a wavelet deconstruction technique, which is also used to characterize spiral structure \citep[e.g.,][]{patrikeev06, frick16}. 
This may even suggest that any differences between these observed features may be artificial.  Several galaxies in the \citet{kalita24} sample targeting clumps have an unmistakable spiral morphology, and some galaxies that we classify as spirals can certainly be described as clumpy.  Many of the clumps observed by CANDELS even appear organized along spiral arms in systems where there are $>$ 10 identified clumps \citep{guo15}.  Now with JWST, those clumps often appear elongated.   
These results may point to a physical link between clumps and spirals, with spiral formation a necessary bottleneck to the formation of clumps that are like the tip of the iceberg, for example.  
Another interpretation is that some detected clumps merely trace the brightest parts of spiral arms.  Whether such clumps are the true clumps of clump formation theory will rely on demonstrating that they are self-gravitating \citep{genzel11} and able to decouple from the arms to migrate to the center.  

\section{Summary and Conclusions}

In this paper we reconsider the origin of the observed distinction between late-type galaxies (dynamically cold, star-forming disks with extended stellar mass profiles) and early-type galaxies (hot, quiescent, and centrally concentrated). Our proposed evolutionary scenario places emphasis on the torques exerted by spiral arms as early as $z\sim 2$ (and even before), but takes as its starting point (Appendix \ref{sec:Norb}) the fiducial disk formation model \citep{fall80, mo98, somerville08, dutton11}, which connects galaxy properties to their dark matter halos.  That model suggests that halo growth leads to galactic disk growth through accretion and the conservation of angular momentum, consistent with observed late-type galaxy size evolution, which is found to be proportional to the evolution of dark matter virial radii \citep{van-der-wel14, huang17, somerville18, mowla19b}.

As we show in this work, this growth can be significantly and rapidly modulated by outward angular momentum transport produced by early, transient, recurrent spiral patterns, which makes galaxies more compact (Section \ref{sec:appl}; Figures \ref{fig:Revol}, \ref{fig:Revol_sample}) and dynamically hot (Sec.~\ref{sec:heating}), consistent with the properties of early-type galaxies. This compaction phase is then followed by dissipationless growth (dry merging) to the present day \citep[e.g.,][]{bezanson09, van-der-wel09, hopkins09a, newman12, van-dokkum14}.  

Our proposed mechanism, EASE -- Early, Accelerated, Secular Evolution -- joins a number of other proposals for producing compaction.  Originally, mergers between spiral galaxies were thought to form ellipticals \citep[][and decades of subsequent work]{toomre72}. More recently, and more akin to the picture presented in this paper, compaction through instabilities in the gas disk has been suggested by \citet{noguchi99, elmegreen07, bournaud07a, genzel08, dekel09} and others as playing an important role at early cosmic times ($1.5<z<3$). The relevant forces and timescales in that picture are similar in magnitude to those in our spiral compaction picture, but the important distinction is that spirals occur in both stellar and gas disks and can therefore act throughout cosmic time, regardless of whether gas fractions are high or low. The relevance of spiral arms across most of cosmic time is highlighted by the breakthrough observation that large fractions of star-forming galaxies at cosmic noon show strong spiral structure \citep[][and Sec.~\ref{sec:emp_class} in this paper]{guo23, kuhn24, espejo-salcedo25}. Examples are shown in Figure \ref{fig:images}.

It is worth emphasizing that spiral arms has long been recognized for the  
influence they can have on the evolution of galactic disks, producing heating, radial migration, (pseudo-)bulge formation and a general evolution from Sc to Sb to Sa and SaB (\citetalias{lynden-bell72}). In this paper we propose for the first time that 
spirals are also responsible for driving the wholesale transformation of massive galaxies from a late-type morphology to an early-type morphology.  Past estimates of spiral arm torques have also found that they are sufficiently strong to radically alter the structure of a galaxy (e.g. \citet{gnedin95} and \citetalias{binney08}).  However, the  significant rearrangement of the angular momentum distribution predicted in 10-30 orbital periods was taken to argue that the very existence of spiral galaxies implies that spiral structure must be short-lived ($\lesssim 1$ Gyr).  Here we turn that argument around and suggest that the rise of the early-type galaxy population, from cosmic noon ($z\approx 2$) to the present, requires precisely a mechanism like this, capable of  rearranging entire galaxies on time scales that are shorter than the Hubble time. 

To support our proposal, in this paper we examine the emergence and influence of the transient nature of spirals.  In the canonical framework of steady spirals, it remains unclear how global changes in the disk density profile can be achieved, given that angular momentum exchanges are generally limited to narrow regions around resonances (corotation and the ILR; \citetalias{lynden-bell72}).  
Such steady spirals are disfavored by modern theory and simulations, which show that transient spirals may be a more natural feature of disks than global stable modes (\citealt{sellwood14,sellwood19,sridhar19}; see also \citealt{lau21, lau21a}). 

As we show in this work (Section \ref{sec:ltrans}), transient spirals are capable of far more widespread changes in angular momentum than steady spirals.  The specific picture of transience that we have in mind emerges from our recent calculation of the characteristic equation for open spirals in the ``Bottom's Dream'' approach (\citetalias{meidt24}).  We show (Section \ref{sec:theory}) that this characteristic equation is able to reproduce a number of behaviors described in the literature \citep{kato71, lynden-bell72,mark74,drury85} and newly underscores how readily transient spirals amplify and decay by leveraging rotation and the self-gravity of the disk.  

We use the patterns of growth predicted by the ``Bottom's Dream'' framework to motivate a heuristic model of spirals (Section \ref{sec:heuristic}) that we then employ for calculating torques and angular momentum transport (Section \ref{sec:torques}).  Our proposed modeling allows the standard calculation of the angular momentum current to perform like the canonical \citetalias{lynden-bell72} torque calculation, but with greater freedom to consider the impact of non-adiabatic spiral growth.  In practice, this allows us to examine the changes to the disk brought about by the torques exerted both steady spirals in the adiabatic limit adopted by \citetalias{lynden-bell72} and transient spirals (Section \ref{sec:diskchange}).   

In the final sections of this paper we applied the theory behind EASE to model the appearance of the early-type galaxy population at cosmic noon as the result of the spiral-driven transformation of late-type galaxies.  To this end, we characterized the spiral population in the range $1<z<2.5$, measuring their abundances and properties (Section \ref{sec:emp}). The majority of star-forming galaxies with stellar masses $M_{\star}>10^{10.5}~M_{\odot}$ have spiral arms, with multiplicities ranging from one to five, with two-armed spirals the most common. The fraction declines with redshift, to about 40\% at $z=2-2.5$ and is weakly mass dependent, with more massive star-forming galaxies more likely to show spiral structure. The ubiquity of spiral structure implies that individually transient spiral features must recur frequently. Their properties (pitch angle and arm contrast, and our measured values of the multiplicity) meanwhile imply that spiral arms can significantly alter their structure of nearly all star-forming galaxies within a few Gyr (Section \ref{sec:appl}). We use our torque formula to predict a reduction in effective radius that matches the observed size distribution of early-type galaxies. 

The framework developed in this paper serves as the foundation to further develop and test the hypothesis of spiral-driven compaction. This starts with more quantitative constraints on the amplitudes and pitch angles of the spiral patterns. With this, prescriptions for compaction rates and how they depend on disk properties can be implemented in galaxy formation models \citep[e.g.,][]{cole00, somerville08, lacey11}
that otherwise invoke disk instabilities in relation to a global $Q$ threshold.  In this work (and \citetalias{meidt24}), on the other hand, the non-axisymmetric structures capable of reshaping disks are transient features that emerge under the continuous action of self-gravity and rotation.  Their growth and decay is an easier, gentler process that does not require a dramatic crossing of the Toomre threshold.  Open, transient spirals even characteristically appear when $1<Q<6$.  This behavior can explain the ubiquity of spirals at all epochs and underscores the importance of a process like EASE across cosmic time.

\begin{acknowledgements}
    A. vdW. and S.E.M. would like to express their gratitude to M. Martorano for laying the groundwork that made the analysis of the observational data in this work possible.  
    A. vdW. would like to thank S. Faber and S.E.M. would like to thank B. Elmegreen for insightful conversation.  
    S.E.M. also acknowledges many stimulating discussions during the IAS `Astrophysical Waves' meeting, including those with C. Hamilton, U. Banik, J. Binney, J. Sellwood and S. Tremaine.
         This work is based on observations made with the NASA/ESA/CSA James Webb Space Telescope. The data were obtained from the Mikulski Archive for Space Telescopes at the Space Telescope Science Institute, which is operated by the Association of Universities for Research in Astronomy, Inc., under NASA contract NAS 5–03127 for JWST. The specific observations analyzed can be accessed via doi \url{10.17909/g3nt-a370}. These observations are associated with programs ERS \#1324, 1345, and 1355; ERO \#2736; GO \#1837 and 2822; GTO \#2738; and COM \#1063. The authors acknowledge the teams and PIs for developing their observing program with a zero-exclusive-access period.
        The data products presented herein were retrieved from the Dawn JWST Archive (DJA). DJA is an initiative of the Cosmic Dawn Center (DAWN), which is funded by the Danish National Research Foundation under grant DNRF140. 

\end{acknowledgements}

\bibliographystyle{aa} 
\bibliography{aanda_aug28_cleaned.bib}

\begin{appendix}
\section{Relation to \citetalias{lynden-bell72}'s donkey effect}\label{sec:donkey}
In \citetalias{meidt24} we related the resonant amplification associated with the short-wave regime to the donkey effect, calculating the motion of orbiting stars in the rotating spiral frame and showing how donkey behavior emerges from different conditions and spiral parameters. Prototypical donkey behavior consists of libration\footnote{Libration refers to the bounded, oscillatory motion in a region of phase space near corotation.} at the spiral potential maximum and motion preferentially directed away from potential minimum.  For large $k$, the spiral forcing is central to the libration (see Eq.~59) in \citetalias{meidt24}) but as $k$ decreases, the rotational properties of disk become the determining factor (see Eq.~55 in \citetalias{meidt24}).  

Even in the long-wave regime, though, spiral forcing remains an influential factor.  In \citetalias{meidt24} we related spiral forcing to a decay that tightens libration around the spiral potential minimum.  Here we will recast that decay in terms of a phase that extends the libration area.  

Consider, first, Eq.~61 of \citetalias{meidt24} (or see \citetalias{binney08}), which highlights motion specifically at the potential extrema, where $\cos {m\phi_0}=\pm 1$ and $\sin {m\phi_0}=0$ with $m\phi_0=m(\Omega-\Omega_p)t$ in terms of the pattern speed $\Omega_p=\omega/m$ of an $m$-armed spiral perturbation.  That particular expression represents the small angle limit of motion along horseshoe orbits and can be obtained by expanding the azimuthal and radial spiral forces in the limit of small radial and azimuthal departures from circular motion ($R_1$ and $\Phi_1$, respectively), using that $\sin(kR_1 +m\Phi_1+m\Phi_0)\approx (kR_1 +m\Phi_1)\cos{m\phi_0}+\sin{m\phi_0}$.  To obtain a wider, less conservative view of the motion in the neighborhood of corotation,  small $\sin$ terms in both the azimuthal and radial forces must be retained.  

We omit a detailed description of the setup used to examine libration motion and refer to \citetalias{meidt24}.  Briefly, we insert a spiral potential perturbation $\Phi_a$ (or density perturbation $\rho_a$) with wavenumber $k$ and multiplicity $m$ into a differentially rotating disk with density $\rho_0$ where the angular rotation speed is $\Omega$.  Shifting into the rotating spiral frame, the radial and azimuthal equations of motion at radius $R_0$ are as given in BT08 or \citetalias{meidt24}. To obtain an expression for the azimuthal acceleration, we first take the time derivative of the perturbed radial equation of motion that describes radial excursions from the circular orbit at $R_0$ and find 
\begin{eqnarray}
&(\kappa_0^2~-~4\Omega_0^2)&\Delta_s\dot{R}_1=2R_0\Omega_0\ddot{\phi_1}\\
&-k\chi_a&\left(\cos(m\phi_0)+(kR_1+m\phi_1)\sin(m\phi_0)\right)m(\Omega-\Omega_p)\nonumber\label{eq:phi1dotderiv}
\end{eqnarray}
where
\begin{equation}
\chi_a=\Phi_a+\sigma^2\frac{\rho_a}{\rho_0}
\end{equation}
including a pressure term associated with velocity dispersion $\sigma^2$.

Then we substitute the resulting expression for $\dot{R}_1$ into the azimuthal equation of motion,  
finding
\begin{eqnarray}
\ddot\phi_1&=&\left(\frac{\kappa_0^2-4\Omega_0^2}{\kappa^2}\right)\frac{m^2}{R_0^2}\chi_a \phi_1\cos{(m\phi_0)}\nonumber\\
&+&\left(\frac{2m\Omega_0(\Omega_0-\Omega_p)}{\kappa^2}\right)\frac{km}{R_0}\chi_a \phi_1\sin{(m\phi_0)}\nonumber\\
&+&\left(\frac{2m\Omega_0(\Omega_0-\Omega_p)}{\kappa^2}\right)\frac{k}{R_0}\chi_a \cos{(m\phi_0)}\nonumber\\
&+&\left(\frac{\kappa_0^2-4\Omega_0^2}{\kappa^2}\right)\frac{m}{R_0^2}\chi_a \cos{(m\phi_0)}.
\label{eq:kmed}
\end{eqnarray}

The first two terms in Eq.~\ref{eq:kmed} represent the libration motion characteristic of the donkey effect, which capitalizes on the behavior of $\kappa^2-4\Omega^2$, which is typically negative in galactic disks (\citetalias{lynden-bell72},\citetalias{binney08}).  Here the radial forcing contributes a term proportional to $\phi_1 \sin{m\phi_0}$, as opposed to the decay term proportional to $\dot{\phi}_1 \cos{m\phi_0}$ in \citetalias{meidt24}.  This makes it clearer that the radial term contributes an azimuthal offset to the cosine term, thus allowing for libration to continue beyond precisely the potential minimum.  In the limit of small $t$, we write the first two terms in Eq.~\ref{eq:kmed} 
as 
\begin{equation}
    \frac{\kappa^2-4\Omega^2}{\kappa^2}\frac{m^2}{R_0^2}\chi_a\cos{(m\phi_0+\phi_{\rm NR})}\label{eq:donkey}
\end{equation}
where
\begin{equation}
\phi_{\rm NR}=\frac{2\Omega}{\tan i_p}\frac{m(\Omega-\Omega_p)}{\Delta}
\end{equation}
where $\tan i_p=m/(kR)$. 
The larger $k$ and $m$, the stronger the potential perturbation and the larger the zone around corotation the libration can extend. 

Likewise, the stronger the potential, the wider is the zone in which orbiting stars or gas parcels preferentially sit on the `downstream' side of the spiral potential minimum, ultimately allowing the wave to grow, as described by \citetalias{lynden-bell72}, across a much larger area.  

The importance of donkey behavior is highlighted (with some algebra) in the context of the ``Bottom's Dream'' characteristic equation in the bottom two lines in Eq.~\ref{eq:longwaveterms}.  Besides a dependence on the factor $\kappa^2-4\Omega^2$, now there is an additional demand primarily placed on the disk density distribution through the gradient in the angular momentum per unit area, $\mathcal{L}$, which is needed in order to insure that sufficient numbers of stars are available on either side of corotation to complete significant wave-amplifying donkey behavior.  Since the third term on the right is mostly always negative in galactic disks, then for the long-wave term to remain negative and yield growth inside corotation, the disk density must fall off rapidly with radius (increase rapidly towards smaller radius). Even moving away from corotation where libration and the donkey effect weakens, wave growth can be stimulated provided that enough stars are present moving inwards away from corotation.   

The net result, according to the solutions to the characteristic equation in Eq.~\ref{eq:long}, is that wave growth either occurs in the form of an island around and beyond the corotation radius for a rigidly rotating wave or as extended shearing material pattern.  In both cases, eventually the growth must cease once the forcing exerted by the spiral no longer supports donkey behavior (because the spiral's amplitude variation exceeds its phase variation) and the wave moves to the decaying wave solution at that location.  We see this here through the $s_0^2$ factor in Eq.~\ref{eq:s0sqlong}, which is the sum of the gravitational and pressure forces.  Growth proceeds as long as the gravitational force dominates but will cease and turn to decay when the pressure force overwhelms gravity.  Since growth increases the $T_r$ factor in the expression for the perturbed potential in Eq.~\ref{eq:poissonFull}, eventually growth shuts off either when the amplitude grows in such a way that $T_r>k$ (for a rigidly rotating wave) or when shear rotates $k$ so that it exceeds $k_J$ (for a shearing material pattern).

\section{Number of Galactic Rotations per Redshift Interval}
\label{sec:Norb}

In Sec.~\ref{sec:ltrans} we describe the change in angular momentum per orbital time. Here we ask how many revolutions disks have gone through as a function of cosmic time.
In standard disk formation theory, galaxies derive their sizes and rotational velocities from their dark matter halos. The halos' sizes and velocities, in turn, are defined with respect to the cosmologically average density. These connections -- between galaxies, halos and cosmology -- are used to predict how galaxy sizes evolve as a function of cosmic time, with reasonable success.  

In essence, the density evolution of the Universe predicts the velocities and sizes of galaxies, that is, their orbital time scales across cosmic time.  The purpose of this section is to estimate the number of galactic revolutions or rotations $N_{\mathrm{rot}}$ as a function of redshift. Such a calculation supersedes the common estimation that $N_{\rm rot}$ is given by the current rotational period (e.g., 200 Myr for the Milky Way) and an age (e.g., 10 Gyr), resulting in $\approx 50$ revolutions. Instead, we will show that disks already had $\approx$50 revolutions between $z=3$ and $z=1$.

First we will derive the orbital time scale for halos as a function of redshift. Then we will convert this to a redshift-dependent orbital time scale for galaxies, defined by their circular velocity at the effective (half-light) radius, and based on theoretically and empirically supported assumptions for the relationship between halo and galaxy properties. This time scale turns out to be independent of mass, and a simple function of redshift (approximately, $\propto \ln(1+z)$), resulting in a similar number of orbits ($\approx 50$) between $z=3$ and $z=1$, and $z=1$ and the present day.

\subsection{The halo orbital time} \label{sec:thalo}
We define the halo orbital time, $T_{\mathrm{orb}}$, as the time for a test particle to complete a circular orbit at the virial radius of a spherical halo.
This quantity 
is closely related to the age of the Universe $t$ and the Hubble parameter $H(t)$. To see this, and analytically derive the relevant time scales, let us first consider a matter-only Universe with $\Omega_{\mathrm{m}}=1$, an Einstein-De Sitter (EdS) Universe with critical density $\rho_{\mathrm{cr}}(t)$:
\begin{equation}H(t) = \sqrt{\frac{8 \pi G \rho_{\mathrm{cr}}(t)}{3}}.\label{eq:TH}\end{equation} 
For this model, the analytical solution relates age to expansion rate as
\begin{equation}t = \frac{2}{3H(t)}.\label{eq:t}\end{equation}

Halos are usually defined as regions with a certain average overdensity $\Delta$ with respect to the cosmic mean density $\rho(t)$ (where, for EdS, $\rho(t)=\rho_{\mathrm{cr}}(t)$). A common choice is $\Delta = 200$, which is an approximation of the analytically derived value $\Delta = 18\pi^2 \approx 178$, the average density within the virial radius of a collapsed top-hat overdensity assuming spherical symmetry and an EdS cosmology \citep{peebles80}. This theoretical value for $\Delta$ is independent of time and the amplitude and size of the initial overdensity (as long as it has collapsed and virialized to form a halo). The implication is that, at a given time $t$, all halos have the same average overdensity within their virial radii, independent of mass.
 The corollary is that the orbital time, which is set by the density under the assumption of spherical symmetry, is independent of halo mass. 
To quantify this, we write the circular velocity of a spherical halo in terms of mass and radius: 
\begin{equation}V^2_{\mathrm{c}}(t) = \frac{GM_{\mathrm{vir}}}{R_{\mathrm{vir}}} = \frac{4\pi G \Delta \rho_{\mathrm{cr}}(t)}{3} R_{\mathrm{vir}}^2.\label{eq:V2}\end{equation}
Inserting $\Delta = 18\pi^2$ and substituting Eq.~\ref{eq:t} and $\rho_{\mathrm{cr}}$ from Eq.~\ref{eq:TH} we find that the halo orbital time is equal to $t$, the age of the Universe, i.e. 
\begin{equation}T_{\mathrm{orb}}(t)\equiv \frac{2\pi R_{\mathrm{vir}}}{V_{\mathrm{c}}}= t.\label{eq:Torb}\end{equation}
As anticipated, $M_{\mathrm{vir}}$ and all other halo-dependencies have dropped out, resulting in an orbital time at the halo virial radius that equals the age of the Universe.  At fixed $t$, this orbital time is independent of halo mass.\footnote{Combining Eqs.~\ref{eq:V2} and ~\ref{eq:Torb} produces $R_{\rm vir} \propto H^{-2/3}$ for fixed $M_{\rm vir}$. This proportionality underpins the theoretical explanation of the observed size evolution of disk galaxies at fixed mass, which we will use in Sec.~\ref{sec:appl}.}

We note that, if $T_{\mathrm{orb}}$ were constant, then the number of orbital times in a time $T$ is $N_{\mathrm{orb}}=T/T_{\mathrm{orb}}$. But since $T_{\mathrm{orb}}$ evolves with time, in each infinitesimal time interval $dt$, the (infinitesimal) number of orbital times is a function of $t$, or
\begin{equation} dN_{\mathrm{orb}}(t) = \frac{dt}{T_{\mathrm{orb}}(t)} = \frac{dt}{t}.\label{eq:dNorb}\end{equation}
The number of orbits over an arbitrary time interval follows after integration over $t$, 
\begin{equation}N_{\mathrm{orb}} (t)= \int dN_{\mathrm{orb}}(t) = \int{\frac{dt}{t}} \propto \ln{t}.\label{eq:Norbt}\end{equation}

To make this useful for interpreting high-redshift observations we change variables from $t$ to redshift $z$. For the adopted EdS cosmology, the relation between time and redshift and its derivative have analytical expressions, 
\begin{equation} t = (1+z)^{-3/2}~;~z=t^{-2/3}-1\end{equation}
and
\begin{equation}\frac{dt}{dz} = -\frac{3}{2}(1+z)^{-5/2}.\end{equation}
Substituting this in to Eq.~\ref{eq:Norbt}, we obtain the number of orbital times since redshift $z$ (between time $t(z)$ and the present, $t_0$) for EdS cosmology: 
\begin{equation}
    N_{\mathrm{orb}}(<z) = \int_{t}^{t_0}  {\frac{dt}{t}} = 
    - \int_{z}^{0}  {\frac{3}{2}\frac{dz}{1+z}} = \frac{3}{2}\ln{(1+z)}.
    \label{eq:Norb}
\end{equation}

What is the relevance of Eq.~\ref{eq:Norb} given the differences between the EdS cosmology adopted so far and $\Lambda$CDM? For $\Lambda$CDM the relation between $z$ and $t$ is different, the density $\rho<\rho_{\mathrm{cr}}$ and depends on time, and so does $\Delta_{\Lambda CDM}$ \citep{bryan98}. Moreover, none of the expressions for $H$, $T_{\mathrm{orb}}$, $t$, and $N_{\mathrm{orb}}$ have analytical analogues for $\Lambda$CDM. Implementing numerical solutions from \textsc{astropy} we have calculated the difference in the number of orbital times before a given redshift, finding the following approximation (accurate to better than 0.2\% for all $z$):

\begin{equation}
    \begin{split}
     (~N_{\rm{orb,\Lambda CDM}} - N_{\mathrm{orb,EdS}}~)~(>z) = - 0.179(1+z)^{-2.5}
    \end{split}\label{eq:dNL}
\end{equation}
which shows that $N_{\rm{orb}}$ is reduced in $\Lambda$CDM compared to an EdS cosmology.  Fewer orbital times per redshift interval are due to longer orbital times at lower halo densities ($\Delta_{\Lambda\mathrm{CDM}} < 18\pi^2$). The increase in time per redshift interval counteracts, but does not entirely compensate for this decrease. The net effect is that, at the present day, there would be 25\% fewer orbital times per incremental increase in scale factor $|\delta a|\approx |\delta z|$ in $\Lambda$CDM vs. EdS cosmology.  As illustrated by the $z$ dependence of Eq.~\ref{eq:dNL}, though, this is mostly a late-time phenomenon given the relatively recent dominance of $\Lambda$. By $z=1$ this difference is 6\% and by $z=2$ it is only 2\%.  This underscores the understanding that, at $z>1$, which is most relevant for this paper, $\Lambda$CDM behaves much like an EdS Universe. Note that Eq.~\ref{eq:Norb} is the integral from $z=0$ to some redshift $z$, whereas it is more natural to consider the dependence on cosmology in the time direction (as in Eq.~\ref{eq:dNL}. This is because the two cosmologies are essentially the same at high $z$, and a significant divergence only arises at late times, $z<1$.

\subsection{Galaxy orbital times} \label{sec:tgal}
Here we derive an estimate of the number of galactic rotations $N_{\mathrm{rot}}$. In a nutshell, we will argue that a reasonable approximation is that galaxy rotational periods (as measured at the effective radius) are 50 times shorter than halo orbital times.  A first step toward justifying this assumption is to consider present-day spiral galaxies with stellar mass with $M_*= 5\times 10^{10}~M_\odot$.  Their typical half-light radius is $R_{\mathrm{gal}}\approx 7$ kpc and rotational velocity $V_{\mathrm{rot}}\approx~200$km s$^{-1}$, so that rotational period at $R_{\mathrm{gal}}$ is
\begin{equation}
T_{\mathrm{rot}} = \frac{2\pi R_{\mathrm{gal}}}{V_{\mathrm{rot}}} \approx 215~\mathrm{Myr}.
\end{equation}
This is $\approx 44$ times shorter than the halo  $T_{\mathrm{orb}}$ derived above (assuming $H_0 = 70~\mathrm{km~s}^{-1}~\mathrm{Mpc}^{-1}$).
This value for $T_{\mathrm{rot}}$ is to first order independent of galaxy mass, because the stellar mass Tully-Fisher and stellar mass - size relations for disk galaxies have approximately the same slope:
\begin{equation}V_{\mathrm{rot}}\propto M^{1/4}~~;~~R_{\mathrm{gal}}\propto M^{1/4}\end{equation}
so that $T_{\mathrm{rot}}\propto R_{\mathrm{gal}}/V_{\mathrm{rot}}\propto M^{0}$. Recall that an analogous lack of mass dependence was obtained for halo orbital times (Eq.~\ref{eq:Torb}).

A similar conclusion can be reached via an alternative, if simpler, route. \citet{kravtsov13, somerville18, mowla19b} showed that the average effective radius of galaxies (at fixed stellar mass) scales linearly with the virial radius of their dark matter halos, with $R_{\mathrm{gal}} \approx 0.02 R_{\mathrm{vir}}$. Supposing a flat rotation curve with rotational velocity $V_{\mathrm{rot}}$ taken to be equal to the circular velocity of the halo, i.e. $V_{\mathrm{rot}} = V_{\mathrm{c}}$, this implies that $T_{\mathrm{rot}}$ is 50 times shorter than $T_{\mathrm{orb}}$.  With this rather simple assumption, we obtain a result that is quite similar to the previous estimate based on observed rotational periods. 

As for halos, the evolution of these galaxy scaling relations with cosmic time is important for estimating the number of orbital times or rotations as a function of time or redshift. The result that $R_{\mathrm{gal}} \approx 0.02 R_{\mathrm{vir}}$ extends to $z\approx 2$ \citep{somerville18, mowla19b}. Ideally, we would combine this with an estimate for $V_{\mathrm{rot}} / V_{\mathrm{c}}$ that applies at higher redshift.  However, a direct estimate of $V_{\mathrm{rot}} / V_{\mathrm{c}}$ as a function of redshift involves several ingredients -- the evolution of the stellar mass Tully-Fisher relation \citep[e.g.,][]{kassin07, dutton11, ubler17, mercier22} and the evolution of the stellar-mass halo-mass relation \citep[e.g.,][]{moster10, behroozi10, rodriguez-puebla17, moster18, behroozi19} -- and is beyond the scope of this paper.  Here we note that several authors \citep[e.g.,][]{burkert16, marasco19} showed that the angular momentum of observed $z\approx 1-2$ disks is in approximate agreement with the theoretical expectations.
We thus continue assuming that  $V_{\mathrm{rot}} = V_{\mathrm{c}}$ hold for disks across cosmic time.  

The assumption $V_{\mathrm{rot}} = V_{\mathrm{c}}$ is a common \textit{ansatz} in standard disk formation theory  \citep{fall80, mo98, van-den-bosch00, somerville08}. 
That theory predicts $R_{\mathrm{gal}} \approx 0.02 R_{\mathrm{vir}}$, implying that galaxies obtain their angular momentum from the halo spin $\lambda$. For example, \citet{mo98} show that in the simplest model, $R_{\mathrm{gal}}=\lambda R_{\mathrm{vir}}/\sqrt{2}$. Numerical simulations show that on average $\lambda\approx 0.04$, so that the predicted
$R_{\mathrm{gal}} / R_{\mathrm{vir}}$ is not far from our approximation. Note that because $R_{\rm vir}\propto H^{-2/3}$ at fixed halo mass (see above) we also have that $R_{\rm gal}\propto H^{-2/3}$ at fixed galaxy mass, in agreement with the observed size evolution of disk galaxies \citep{van-der-wel14}.

In conclusion, taking all empirical and theoretical evidence together, it seems more than reasonable to assume $V_{\mathrm{rot}} = V_{\mathrm{c}}$ and $R_{\mathrm{gal}} = 0.02 R_{\mathrm{vir}}$ for converting halo orbital times to the rotational periods of galactic disks at their effective radius. Since the effective radius is the median radius of stars, the derived rotational period should be seen as the median rotational period of stars. 

We emphasize that, given the equivalence of halo orbital time and age of the Universe for the EdS cosmology (Sec.~\ref{sec:thalo}), then the result of the above assumptions is that the galaxy rotational period (at the effective radius) is always 2\% of the age of the Universe.  Thus, we can obtain the number of revolutions at $R_{\mathrm{gal}}$ (the half-light radius) simply by multiplying Eq.~\ref{eq:Norb} by a factor 50. 
For an Einstein-De Sitter cosmology this is:

\begin{equation}
    N_{\mathrm{rot}}(<z) = 75~\ln{({1+z)} }
\end{equation}

or
\begin{equation}
    N_{\mathrm{rot}}~(z_1<z<z_2) = 75~\ln{\left ( \frac{1+z_2}{1+z_1} \right)}
    \label{eq:Nrotz}.
\end{equation}
To adjust this to $\Lambda$CDM, we use the same adjustment as in Eq.~\ref{eq:dNL}, modulo the factor 50, i.e. 
\begin{equation}
     (~N_{\rm{rot,\Lambda CDM}} - N_{\mathrm{rot,EdS}}~)~(>z) = - 8.95(1+z)^{-2.5}.
    \label{eq:Nadj}
\end{equation}
This goes to zero for $z\gg 0$, but, as noted earlier in Section \ref{sec:thalo}, the steep redshift dependence implies that $N_{\rm{rot,\Lambda CDM}}$ deviates from $N_{\rm{rot,EdS}}$ by less than 1\% for $z>1$. In conclusion, for estimating $N_{\rm rot}$ in any redshift interval at $z>1$ we recommend using Eq.~\ref{eq:Nrotz}. For $z<1$ we recommend adjusting $N_{\rm rot}$ according to Eq.~\ref{eq:Nadj}.

For the redshift interval that we examine in this paper ($1<z<2.5$), we have $N_{\mathrm{rot}}\approx 41$ for $\Lambda$CDM and $N_{\mathrm{rot}}\approx 42$ for EdS, a negligible difference given the underlying assumptions and uncertainties regarding halo circular velocities and their conversion to galactic orbital times. 

Notably, $N_{\mathrm{rot}}(0<z<1) \approx N_{\mathrm{rot}}(1<z<3) \approx 50$, regardless of cosmology (for EdS, $N_{\mathrm{rot}}(0<z<1) \equiv N_{\mathrm{rot}}(1<z<3)$), i.e.  
there are as many galactic revolutions at late times (between $z=1$ and the present)  as there were at "cosmic noon" (between $z=3$ and $z=1$).  This leads to the conclusion that  processes typically referred to as secular in present-day galaxies are equally impactful at the heyday of star- and galaxy formation.  If morphological features, such as bars and spiral arms, were as prevalent then as they are now, shorter orbital times imply that they can even more rapidly heat disks and redistribute angular momentum than today. 

\end{appendix}

\end{document}